\def\complexNumbers{\mathbb{C}}
\def\realNumbers{\mathbb{R}}
\def\constante{{\rm e}}

\def\EDreceivedsignal[#1][#2]{r^{#1,{#2}}_{\rm ED}}
\def\ESreceivedsignal[#1][#2]{r^{#1,{#2}}_{\rm ES}}

\def\receivedPowerOfED[#1][#2]{P^{#1,#2}_{\rm ED}}
\def\receivedPowerOfES[#1][#2]{P^{#1,#2}_{\rm ES}}
\def\constantPowerOfED{}
\def\constantPowerOfES{}
\def\numberofconnectedED{\numberOfEdgeDevices_{\rm c}}
\def\numberofconnectedES{\numberOfEdgeServers_{\rm c}}
\def\probability[#1]{\mathbb{P}\left({#1}\right)}
\def\binomialRandomVariable[#1][#2]{B(#1,#2)}
\def\indicatorFunction[#1]{\mathbb{I}\left[{#1}\right]}
\def\indexES{s}
\def\indexED{k}
\def\DLchannel[#1][#2]{h^{#1,{#2}}_{\rm DL}}
\def\ULchannel[#1][#2]{h^{#1,{#2}}_{\rm UL}}
\def\noiseEDplus[#1][#2]{\omega^{#1,{#2}}_{\rm ED}}
\def\noiseESplus[#1][#2]{\omega^{#1,{#2}}_{\rm ES}}
\def\psiforp[#1][#2]{\Delta^{#1,{#2}}_{\rm ED}}
\def\deltaofr[#1][#2]{\Delta^{#1,{#2}}_{\rm ES}}
\def\signofdelta[#1][#2]{\textbf{v}^{#1,{#2}}_{\rm ES}}
\def\signofdeltaElement[#1][#2]{v^{#1,{#2}}_{\rm ES}}
\def\signofpsi[#1][#2]{\textbf{v}^{#1,{#2}}_{\rm ED}}
\def\signofpsiElement[#1][#2]{v^{#1,{#2}}_{\rm ED}}
\def\globalgrad[#1]{\textbf{g}^{(#1)}}
\def\indexofconstantvector[#1]{L_{#1}}
\def\constantvector{\textbf{L}}

\def\numberOfEDsWithCorrectChoice{Z}
\def\numberOfESsWithCorrectChoice{Y}
\def\indexGradient{i}
\def\expectationOperator[#1][#2]{{\mathbb{E}_{#2}}\left[#1\right]}
\def\probabilityIncorrect[#1]{P^{\rm{err}}_{#1}}
\def\numberOFEDsForOptionOne{K^{+}_{\indexES}}
\def\numberOFEDsForOptionSecond{K^{-}_{\indexES}}
\def\numberOFESsForOptionOne{S^{+}_{\indexED}}
\def\numberOFESsForOptionSecond{S^{-}_{\indexED}}
\def\correctDecision[#1]{p_{\textrm{z},#1}}
\def\incorrectDecision[#1]{{(1-\correctDecision[\indexGradient])}}
\def\correctDecisionED{p_{\textrm{y},\indexGradient}}
\def\incorrectDecisionED{{(1-\correctDecisionED)}}
\def\indexSubcarrier{m}
\def\indexOFDMSymbol{n}
\def\voteInTime[#1]{n^{#1}}
\def\voteInFrequency[#1]{m^{#1}}
\def\randomSymbolAtSubcarrierED[#1]{s_{\rm ED}^{\indexCommunicationRound,{#1}}}
\def\transmittedSymbolAtSubcarrierED[#1]{t_{\rm ED}^{\indexCommunicationRound,{#1}}}
\def\transmittedSymbolAtSubcarrierES[#1]{t_{\rm ES}^{\indexCommunicationRound,{#1}}}
\def\randomizationSymbolAtSubcarrierED[#1]{s_{\rm ED}^{\indexCommunicationRound,{#1}}}
\def\randomizationSymbolAtSubcarrierES[#1]{s_{\rm ES}^{\indexCommunicationRound,{#1}}}
\def\meanOptionOne{\mu^{+}_{{\rm ED,\indexGradient}}}
\def\meanOptionTwo{\mu^{-}_{{\rm ED,\indexGradient}}}
\def\meanOptionOneES{\mu^{+}_{{\rm ES,\indexGradient}}}
\def\meanOptionTwoES{\mu^{-}_{{\rm ES,\indexGradient}}}
\def\noiseVariance{\sigma_{\rm ES}^2}
\def\noiseVarianced{\sigma_{\rm ED}^2}
\def\batchSize{n_{\rm b}}
\def\varianceBound{{\bm \sigma}}
\def\indexCommunicationRound{t}
\def\communicationRounds{T}
\def\learningRate{\eta}
\def\modelParametersOptimal{{\textbf{w}_{\indexED}^{*}}}

\def\modelParameters{\textbf{w}_{\indexED}}
\def\completeData{\mathcal{G}}
\def\localClasses[#1]{\mathcal{C}_{#1}}
\def\modelParametersAtIterationEle[#1][#2]{w^{(#1)}_{#2}}
\def\datasetBatch[#1]{\mathcal{\tilde{D}}_{#1}}
\def\majorityVote[#1][#2]{\textbf{v}^{#1,{#2}}_{\rm ES}}
\def\GlobalusercentricLF[#1][#2]{F_{#1}(#2)}
\def\localdataset[#1]{\mathcal{G}_{#1}}

\def\localGradientSign[#1][#2]{{\bar{\textbf{g}}}_{#1}^{(#2)}}
\def\localGradientSignElement[#1][#2]{{\bar{g}}_{#1}^{(#2)}}
\def\lossFunction[#1]{{F}_{#1}}
\def\globalGradient[#1]{\hat{\textbf{{g}}}^{(#1)}}
\def\localGradient[#1][#2]{\textbf{g}_{#1}^{(#2)}}
\def\localGradientNoIndex[#1]{{\textbf{g}}_{#1}}
\def\ElementsOfGradient[#1][#2]{{g}_{#1}^{(#2)}}
\def\ElementsOfGradientNoindex[#1]{{g}_{#1}}
\def\localGradientStat[#1][#2]{{\tilde{\textbf{g}}}_{#1}^{(#2)}}
\def\localGradientStatElement[#1][#2]{{\tilde{g}^{(#2)}}_{#1}}
\def\localGradientStatNoIndex[#1]{{\tilde{\textbf{g}}}_{#1}}
\def\localGradientStatNoIndexElement[#1]{{\tilde{g}}_{#1}}
\def\SignofUsercentricElement[#1][#2]{\hat{g}^{(#2)}_{#1}}
\def\symbolEnergy{E_{\rm s}}

\def\numberOfEdgeServers{S}
\def\numberOfEdgeDevices{K}
\def\dataset[#1]{\mathcal{D}_{#1}}

\def\modelParametersAtIteration[#1][#2]{\textbf{w}^{(#1)}_{#2}}

\def\numberOfModelParametersED[#1]{Q_{#1}}
\def\sampleData[#1]{{\textrm{\textbf{x}}}_{#1}}
\def\indexSampleData{{\ell}}
\def\sampleLabel[#1]{{y}_{#1}}
\def\aggregatedGradientSigns[#1]{\textbf{y}^{(#1)}}
\def\aggregatedGradientSignsEle[#1][#2]{{y}^{(#1)}_{#2}}
\def\indexGradient{i}
\def\globalGradientSignEle[#1][#2]{{v}^{(#1)}_{#2}}
\def\globalGradientSign[#1]{\textbf{v}^{(#1)}}
\def\majorityVoteAtIteration[#1]{\textbf{v}^{(#1)}}

\def\lossFunctionSampleLocal[#1][#2]{f(#1)}
\def\lossFunctionSample[#1]{f(#1)}
\def\lossFunctionLocal[#1][#2]{F_{#1}(#2)}
\def\lossFunctionGlobal[#1][#2]{F_{#1}(#2)}
\def\lossFunctionGlobal[#1]{F(#1)}

\def\neuralNetworkAtED[#1]{\mathcal{N}_{#1}}

\def\updateRule[#1][#2]{U_{#1}(#2)}

\def\symbolVector[#1]{\textbf{\textrm{d}}_{#1}}
\def\symbolVectorEstimate[#1]{\tilde{\textbf{\textrm{{d}}}}_{#1}}
\def\receivedVector[#1]{\textbf{\textrm{r}}_{#1}}
\def\noiseVector[#1]{\textbf{\textrm{n}}_{#1}}
\def\transmittedVector[#1]{\textbf{\textrm{t}}_{#1}}
\def\idftMatrix[#1]{\textbf{\textrm{F}}_{#1}^{\rm H}}
\def\dftMatrix[#1]{\textbf{\textrm{F}}_{#1}}
\def\transformPrecoder[#1]{\textbf{\textrm{T}}_{#1}}
\def\transformDecoder[#1]{\textbf{\textrm{T}}_{#1}^{\rm -1}}

\def\channelMatrix[#1]{\textbf{\textrm{H}}_{#1}}

\def\timeDomainOFDM[#1]{x(#1)}

\def\numberOfActiveSubcarriers{M}

\def\dataSymbols[#1]{d_{#1}}

\def\channelAtSubcarrier[#1]{h_{#1}}
\def\noiseAtSubcarrier[#1]{n_{#1}}
\def\numberOfOFDMSymbols{N}
\def\asymbolFromED[#1]{d_{#1}}
\def\exponentialIntegral[#1]{{\rm Ei}(#1)}
\def\tciFactor[#1]{p_{#1}}
\def\encoder[#1]{\psi_#1}
\def\symbolEnergy{E_{\rm s}}

\def\syncError{T_{\rm sync}}

\def\variableForGeoemtricMean{{\rm \bf z}}
\def\variableForGeoemtricMeanIteration[#1]{\variableForGeoemtricMean^{#1}}

\def\weightDistance[#1]{\alpha_{#1}}
\def\medianDistance[#1]{\beta_{#1}}

\def\indexArea{a}
\def\referenceDistanceUplink{r_{\rm UL}}
\def\referenceDistanceDownlink{r_{\rm DL}}
\def\distanceED[#1]{r_{#1}}
\def\pathlossExponent{\alpha}
\def\modelParametersEle[#1]{{w}_{#1}}

\def\nonnegativeConstants{\textbf{L}}
\def\nonnegativeConstantsEle[#1]{L_{#1}}
\def\varianceBoundEle[#1]{\sigma_{#1}}
\def\lossFunctionGlobalMinimum{F^*}

\def\globalGradientElementNoIndex[#1]{{g_{#1}}}
\def\globalGradient[#1]{{\textbf{{g}}}^{(#1)}}
\def\globalGradientElement[#1][#2]{{{g}}^{(#1)}_{#2}}

\documentclass[conference, comsoc]{IEEEtran}
\usepackage{multicol}
\usepackage{lipsum}
\usepackage{mathtools}
\usepackage{amsthm}
\usepackage{acronym}
\usepackage[bookmarksopen=true]{hyperref}
\usepackage{stfloats}
\usepackage{amsfonts}
\usepackage{acronym}
\usepackage{cite}
\usepackage{multirow}
\usepackage{bm}
\usepackage{amsmath,amssymb}
\usepackage{graphicx}
\usepackage{epstopdf}
\usepackage[table,xcdraw]{xcolor}
\usepackage[caption=false,font=footnotesize]{subfig}
\usepackage[geometry]{ifsym}
\usepackage{array}
\usepackage[utf8]{inputenc}
\usepackage[T1]{fontenc}
\usepackage{algpseudocode}
\usepackage[ruled,vlined]{algorithm2e}

\let\norm\undefined % <-- "Undefine" \norm
\DeclarePairedDelimiter\norm{\lVert}{\rVert}
\newcommand\mydots{\hbox to 1em{.\hss.\hss.}}

\newtheorem{theorem}{Theorem}

\newtheorem{assumption}{Assumption}

\DeclareMathOperator{\sign}{sign}

\makeatletter
\newif\ifAC@uppercase@first%
\def\Aclp#1{\AC@uppercase@firsttrue\aclp{#1}\AC@uppercase@firstfalse}%
\def\AC@aclp#1{%
	\ifcsname fn@#1@PL\endcsname%
	\ifAC@uppercase@first%
	\expandafter\expandafter\expandafter\MakeUppercase\csname fn@#1@PL\endcsname%
	\else%
	\csname fn@#1@PL\endcsname%
	\fi%
	\else%
	\AC@acl{#1}s%
	\fi%
}%
\def\Acp#1{\AC@uppercase@firsttrue\acp{#1}\AC@uppercase@firstfalse}%
\def\AC@acp#1{%
	\ifcsname fn@#1@PL\endcsname%
	\ifAC@uppercase@first%
	\expandafter\expandafter\expandafter\MakeUppercase\csname fn@#1@PL\endcsname%
	\else%
	\csname fn@#1@PL\endcsname%
	\fi%
	\else%
	\AC@ac{#1}s%
	\fi%
}%
\def\Acfp#1{\AC@uppercase@firsttrue\acfp{#1}\AC@uppercase@firstfalse}%
\def\AC@acfp#1{%
	\ifcsname fn@#1@PL\endcsname%
	\ifAC@uppercase@first%
	\expandafter\expandafter\expandafter\MakeUppercase\csname fn@#1@PL\endcsname%
	\else%
	\csname fn@#1@PL\endcsname%
	\fi%
	\else%
	\AC@acf{#1}s%
	\fi%
}%
\def\Acsp#1{\AC@uppercase@firsttrue\acsp{#1}\AC@uppercase@firstfalse}%
\def\AC@acsp#1{%
	\ifcsname fn@#1@PL\endcsname%
	\ifAC@uppercase@first%
	\expandafter\expandafter\expandafter\MakeUppercase\csname fn@#1@PL\endcsname%
	\else%
	\csname fn@#1@PL\endcsname%
	\fi%
	\else%
	\AC@acs{#1}s%
	\fi%
}%
\edef\AC@uppercase@write{\string\ifAC@uppercase@first\string\expandafter\string\MakeUppercase\string\fi\space}%
\def\AC@acrodef#1[#2]#3{%
	\@bsphack%
	\protected@write\@auxout{}{%
		\string\newacro{#1}[#2]{\AC@uppercase@write #3}%
	}\@esphack%
}%
\def\Acl#1{\AC@uppercase@firsttrue\acl{#1}\AC@uppercase@firstfalse}
\def\Acf#1{\AC@uppercase@firsttrue\acf{#1}\AC@uppercase@firstfalse}
\def\Ac#1{\AC@uppercase@firsttrue\ac{#1}\AC@uppercase@firstfalse}
\def\Acs#1{\AC@uppercase@firsttrue\acs{#1}\AC@uppercase@firstfalse}

\acrodef{signSGD}{sign stochastic gradient descend}
\acrodef{MV}{majority vote}

\acrodef{SNR}{signal-to-noise ratio}
\acrodef{RMSE}{root-mean-square error}
\acrodef{OFDM}{orthogonal frequency division multiplexing}
\acrodef{DFT}{discrete Fourier transform}
\acrodef{PSK}{phase-shift keying}
\acrodef{QAM}{quadrature amplitude modulation}
\acrodef{QPSK}{quadrature phase-shift keying}
\acrodef{PMEPR}{peak-to-mean envelope power ratio}
\acrodef{BER}{bit-error ratio}
\acrodef{SNR}{signal-to-noise ratio}
\acrodef{PSD}{power spectral density}
\acrodef{SE}{spectral efficiency}
\acrodef{CP}{cyclic prefix}
\acrodef{AWGN}{additive white Gaussian noise}
\acrodef{CFR}{channel frequency response}
\acrodef{CIR}{channel impulse response}
\acrodef{MMSE}{minimum mean square error}
\acrodef{LMMSE}{linear minimum mean square error}
\acrodef{BPSK}{binary phase shift keying}
\acrodef{BLER}{block-error rate}
\acrodef{ML}{maximum likelihood}
\acrodef{PHY}{physical layer}
\acrodef{PA}{power amplifier}
\acrodef{IDFT}{inverse DFT}
\acrodef{DoF}{degrees-of-freedom}
\acrodef{IoT}{Internet-of-Things}
\acrodef{FDE}{frequency-domain equalization}
\acrodef{FSK}{frequency-shift keying}
\acrodef{FSK-MV}{\ac{FSK}-based \ac{MV}}
\acrodef{PPM}{pulse-position modulation}
\acrodef{PPM-MV}{\ac{PPM}-based \ac{MV}}
\acrodef{RF}{radio-frequency}
\acrodef{IM}{index modulation}
\acrodef{BS}{base station}
\acrodef{MF}{matched filter}

\acrodef{BAA}{broadband analog aggregation}
\acrodef{OBDA}{one-bit broadband digital aggregation}
\acrodef{FEEL}{federated edge learning}
\acrodef{FL}{federated learning}
\acrodef{ED}{edge device}
\acrodef{ES}{edge server}
\acrodef{UL}{uplink}
\acrodef{DL}{downlink}
\acrodef{OAC}[OAC]{over-the-air computation}
\acrodef{TCI}{truncated-channel inversion}
\acrodef{MV}{majority vote}
\acrodef{CNN}{convolution neural network}
\acrodef{ReLU}{rectified-linear unit}
\acrodef{CSI}{channel state information}
\acrodef{PAPR}{peak-to-average power ratio}
\acrodef{iid}[IID]{independent and identically distributed}
\acrodef{5G}{Fifth Generation}
\acrodef{4G}{Fourth Generation}
\acrodef{NR}{New Radio}
\acrodef{LTE}{Long Term Evolution}
\acrodef{RACH}{random-access channel}
\acrodef{DNN}{deep nueral network}
\acrodef{SGD}{stochastic gradient descend}
\acrodef{SGD}{stochastic gradient descend}
\acrodef{signSGD}{sign stochastic gradient descend}

\acrodef{5G}{Fifth Generation}
\acrodef{4G}{Fourth Generation}
\acrodef{NR}{New Radio}
\acrodef{PRACH}{physical random access channel}
\acrodef{PUCCH}{physical uplink control channel}
\acrodef{OFDMA}{orthogonal frequency division multiple access}

\begin{document}

\title{Multi-cell Non-coherent Over-the-Air Computation
for Federated Edge Learning }

% \author{
% 	\IEEEauthorblockN{Mohammad Hassan Adeli}
% 	\IEEEauthorblockA{Electrical  Engineering Department\\
% 		University of South Carolina\\
% 		Columbia, SC, USA\\
% 		Email: madeli@email.sc.edu}
% 		\and
% 	\IEEEauthorblockN{Alphan \c{S}ahin}
% 	\IEEEauthorblockA{Electrical  Engineering Department\\
% 	University of South Carolina\\
% 	Columbia, SC, USA\\
% 	Email: asahin@mailbox.sc.edu}
% }

\author{
	\IEEEauthorblockN{Mohammad Hassan Adeli and Alphan \c{S}ahin}
	\IEEEauthorblockA{Electrical  Engineering Department, University of South Carolina, Columbia, SC, USA\\
		Email: madeli@email.sc.edu, asahin@mailbox.sc.edu}
}

\maketitle

\begin{abstract}
In this paper, we propose a framework where \ac{OAC} occurs in both \ac{UL} and \ac{DL}, sequentially, in a multi-cell environment to address the latency and the scalability issues of \ac{FEEL}.  To eliminate the \ac{CSI} at the \acp{ED} and \acp{ES} and relax the time-synchronization requirement  for the \ac{OAC}, we use a non-coherent computation scheme, i.e., \ac{FSK-MV}. With the proposed framework,  multiple \acp{ES} function as the aggregation nodes in the \ac{UL} and each \ac{ES} determines the \acp{MV} independently. After the \acp{ES} broadcast the detected \acp{MV}, the \acp{ED} determine the sign of the gradient through another \ac{OAC} in the \ac{DL}. Hence, inter-cell interference is exploited for the \ac{OAC}. In this study, we prove the convergence of the non-convex optimization problem for the \ac{FEEL} with the proposed \ac{OAC} framework. We also numerically evaluate the efficacy of the proposed method by comparing the test accuracy in both multi-cell and single-cell scenarios for both homogeneous and heterogeneous data distributions. %We demonstrate  improvements of test accuracy in our \ac{OAC} method in the multi-cell scenario. Moreover, we demonstrate the superiority of our approach by performing training based on only local data in terms of accuracy.
\end{abstract}

% \begin{IEEEkeywords}
% Over-the-air computation
% \end{IEEEkeywords}
\acresetall
\section{Introduction}
\Ac{OAC} refers to the computation of mathematical functions by exploiting the superposition property of wireless multiple-access channel  \cite{Nazer_2007}. It has been initially considered in wireless sensor networks to reduce the latency due to a large number of nodes  \cite{Goldenbaum_2013, Tang_2021_aschnPulse, Chen_2018_AirComp}. Recently, it has been shown that \Ac{OAC} is also a prominent solution to address the latency issue of \ac{FEEL} \cite{gafni2021federated} or distributed training problems in a wireless network \cite{chen2021distributed}. %In, a client-edge-cloud  \ac{FL} architecture is proposed to take advantage of both cloud-based and edge-based \ac{FL} to achieve faster training and lower energy consumption. A hierarchical three-layer \ac{FL} system is proposed which in addition to communication and aggregation steps between clients and the cloud, it includes some \acp{ES} for local aggregation of its client. 
Nevertheless, apart from a few work, e.g.,  \cite{9148862}, \ac{FEEL} with \ac{OAC} is primarily investigated in a single cell in the \ac{UL} although the practical wireless networks often consist of multiple cells. In this study, we address this issue and propose a framework  for \ac{FEEL} based on a non-coherent \ac{OAC} scheme in both \ac{UL} and \ac{DL} in a multi-cell environment.

One of the major challenges in the \ac{OAC} is the  detrimental impact of wireless channels on the coherent symbol superposition. To address this issue, a majority of the state-of-the-art solutions rely on pre-equalization techniques. For instance, in \cite{Guangxu_2020} and \cite{Amiri_2020}, \ac{BAA} over \ac{OFDM} with \ac{TCI} is investigated to obtain unbiased estimates of the weights or gradients. In \cite{Guangxu_2021}, \ac{OBDA}, inspired by distributed training by \ac{MV} with the  \ac{signSGD}~\cite{Bernstein_2018}, is proposed to facilitate the implementation of \ac{FEEL} for a practical wireless system, which also uses \ac{TCI}. %With \ac{OBDA}, the \acp{ED} transmit the signs of the stochastic gradients, i.e., votes and the \ac{ED} obtains the \ac{MV} by calculating the signs of the real and imaginary components of the superposed received symbols on each subcarrierX.
In \cite{Liqun_2021_hier}, instead of \ac{TCI}, the conjugate of the channel is utilized. In \cite{Yang_2020} and \cite{Amiria_2021}, it is assumed that the \ac{CSI} for each \ac{ED} is available at the \ac{ES}. The impact of the channel on \ac{OAC} is mitigated through beamforming techniques. In this paper, to overcome this challenge and requiring \ac{CSI} access at the \acp{ED} and the \acp{ES}, a non-coherent computation scheme is deployed.

The state-of-the-art \ac{OAC} techniques are often suitable for a single cell where the \ac{OAC} occurs in the \ac{UL} due to the pre-equalization. %The main bottleneck for the extensions of these schemes to a multi-cell environment or a \ac{DL} scenario requires consideration of \ac{CSI} at multiple nodes as they rely on coherent computation. 
In addition, pre-equalization techniques require sample-level precise time synchronization, which causes another shortcoming when multiple aggregation nodes exist in a wireless network. In \cite{sahinCommnet_2021} and \cite{sahinWCNC_2022}, non-coherent computation through \ac{FSK-MV} and \ac{PPM-MV} are investigated for \ac{FEEL} in a single cell scenario. The main strategy in these studies is to dedicate two resources where either of the two resources are activated based on the sign of the gradient. The \ac{MV} at the \ac{ES} is detected through an energy detector. Since the information is not encoded in the amplitude or the phase in this strategy, the need for \ac{CSI} at the \acp{ED} and the \ac{ES} are eliminated and the precise time-synchronization requirement is relaxed. Because of these unique features, we consider non-coherent \ac{OAC} in a multi-cell environment. 

In this study, we propose an \ac{OAC} framework where \ac{OAC} occurs in both \ac{UL} and \ac{DL} in a multi-cell environment with \ac{FSK}-based \ac{MV}. As opposed to a single-cell solution, multiple \acp{ES} first detect the \acp{MV} through the \ac{UL} \ac{OAC}. Afterward, each \ac{ED} determines the sign of the gradient by aggregating the \acp{ES}' signals  in the \ac{DL} with another \ac{OAC}. We show the convergence of the non-convex loss function problem for \ac{FEEL} with the proposed scheme and evaluate the proposed framework numerically. We show that the efficacy of the proposed framework by comparing it with a single-cell scenario for both homogeneous and heterogeneous data distributions.

{\em Notation:} %The sets of complex numbers and real numbers are denoted by $\complexNumbers$ and $\realNumbers$, respectively. 
$\expectationOperator[\cdot][]$ is the expectation operation.  
$\indicatorFunction[\cdot]$ is the indicator function. The  function  $\sign(\cdot)$ results in $1$, $-1$, or $\pm1$ at random for a positive, a negative, or a zero-valued argument. 

\section{System Model}
\label{sec:system}
%\subsection{Deployment}
Consider a multi-cell wireless network with $\numberOfEdgeDevices$ \acp{ED} and $\numberOfEdgeServers$ \acp{ES}. We assume that the frequency synchronization in the network is handled through a control mechanism. We consider time synchronization errors among the \acp{ED} (and the \acp{ES}) and the maximum difference between the time of arrivals of the signals at the desired receiver's location is $\syncError$~seconds, where $\syncError$ is equal to the reciprocal to the signal bandwidth. We assume that the \ac{UL} \ac{SNR} at an \ac{ES} is $1/\noiseVariance$ when an \ac{ED} is located at the reference distance $\referenceDistanceUplink$. We then set the received signal power of the $\indexED$th \ac{ED} at the $\indexES$th \ac{ES} as
$\receivedPowerOfED[\indexED][\indexES]={\distanceED[\indexED,\indexES]}^{-\pathlossExponent}/{\referenceDistanceUplink^{-\pathlossExponent}}$, 
where $\distanceED[\indexED,\indexES]$ is the link distance between the $\indexED$th \ac{ED} and the $\indexES$th \ac{ES}, and $\pathlossExponent$ is the path loss exponent. Similarly, we define the \ac{DL} \ac{SNR} at an \ac{ED} is $1/\noiseVarianced$ when the distance between an \ac{ED} and an \ac{ES} is equal to the reference distance $\referenceDistanceDownlink$. We then set the received signal power of the  $\indexES$th \ac{ES}  at the $\indexED$th \ac{ED} as
$\receivedPowerOfES[\indexES][\indexED]={{\distanceED[\indexES,\indexED]}^{-\pathlossExponent}}/{\referenceDistanceDownlink^{-\pathlossExponent}}$.

\subsection{Signal Model in Uplink and Downlink}
In this study, the \acp{ED} in the \ac{UL} and the \acp{ES} in the \ac{DL} access the wireless channel  on the same time-frequency resources simultaneously with $\numberOfOFDMSymbols$ \ac{OFDM} symbols consisting of $\numberOfActiveSubcarriers$ active subcarriers.  We assume that the \ac{CP} duration is larger than  $\syncError$ and the maximum-excess delay of the channel.
Considering independent frequency-selective channels between the \acp{ED} and the \acp{ES}, the superposed symbol on the $\indexSubcarrier$th subcarrier of the $\indexOFDMSymbol$th \ac{OFDM} symbol at the $\indexES$th \ac{ES}  for the $\indexCommunicationRound$th communication  round of \ac{FEEL} can  be written as
\begin{align}
	\ESreceivedsignal[\indexCommunicationRound][{\indexES,\indexSubcarrier,\indexOFDMSymbol}] = \sum_{\indexED=1}^{\numberOfEdgeDevices}\sqrt{\receivedPowerOfED[\indexES][{\indexED}]}\ULchannel[\indexCommunicationRound][\indexES,\indexED,\indexSubcarrier,\indexOFDMSymbol]\transmittedSymbolAtSubcarrierED[\indexED,\indexSubcarrier,\indexOFDMSymbol]+\noiseESplus[\indexCommunicationRound][\indexES,\indexSubcarrier,\indexOFDMSymbol]~,
	\label{eq:symbolOnSubcarrierUplink}
\end{align}
where  $\ULchannel[\indexCommunicationRound][\indexES,\indexED,\indexSubcarrier,\indexOFDMSymbol]\in\complexNumbers$ is the channel coefficient between the $\indexES$th \ac{ES} and the $\indexED$th \ac{ED}, $\transmittedSymbolAtSubcarrierED[\indexED,\indexSubcarrier,\indexOFDMSymbol]\in\complexNumbers$ is the transmitted symbol from the $\indexED$th \ac{ED}, and $\noiseESplus[\indexCommunicationRound][\indexED,\indexSubcarrier,\indexOFDMSymbol]$ is the symmetric \ac{AWGN} with zero mean and the variance $\noiseVariance$ on the $\indexSubcarrier$th subcarrier for $\indexSubcarrier\in\{0,1,\mydots,\numberOfActiveSubcarriers-1\}$ and $\indexOFDMSymbol\in\{0,1,\mydots,\numberOfOFDMSymbols-1\}$. Similarly, the received symbol  on the $\indexSubcarrier$th subcarrier of the $\indexOFDMSymbol$th \ac{OFDM} symbol at the $\indexED$th \ac{ED}  for the $\indexCommunicationRound$th communication  round in the \ac{DL} can be shown as
\begin{align}
   \EDreceivedsignal[\indexCommunicationRound][{\indexED,\indexSubcarrier,\indexOFDMSymbol}]=\sum_{\indexES=1} ^{\numberOfEdgeServers}\sqrt{\receivedPowerOfES[\indexES][\indexED]}\DLchannel[\indexCommunicationRound][{\indexES,\indexED,\indexSubcarrier,\indexOFDMSymbol}]\transmittedSymbolAtSubcarrierES[{\indexES,\indexSubcarrier,\indexOFDMSymbol}]+\noiseEDplus[\indexCommunicationRound][{\indexED,\indexSubcarrier,\indexOFDMSymbol}]~,
   \label{eq:symbolOnSubcarrierDownlink}
\end{align}
where $\DLchannel[\indexCommunicationRound][{\indexES,\indexED,\indexSubcarrier,\indexOFDMSymbol}]\in\complexNumbers$ is the channel coefficient between the $\indexES$th \ac{ES} and the $\indexED$th \ac{ED}, $\transmittedSymbolAtSubcarrierES[{\indexES,\indexSubcarrier,\indexOFDMSymbol}]\in\complexNumbers$ is the transmitted symbol from the $\indexES$th \ac{ES}, and $\noiseEDplus[\indexCommunicationRound][{\indexED,\indexSubcarrier,\indexOFDMSymbol}]$ is the symmetric \ac{AWGN} with zero mean and the variance $\noiseVarianced$ on the $\indexSubcarrier$th subcarrier.
%for $\indexSubcarrier\in\{0,1,\mydots,\numberOfActiveSubcarriers-1\}$ and $\indexOFDMSymbol\in\{0,1,\mydots,\numberOfOFDMSymbols-1\}$.

\subsection{Problem Statement and Learning Model}
\label{subsec:learning}
Let $\modelParametersAtIteration[\indexCommunicationRound][\indexED]\in\realNumbers^{\numberOfModelParametersED[]}$ denote the model parameters at the $\indexED$th \ac{ED} for the $\indexCommunicationRound$th communication round, and $\numberOfModelParametersED[]$ is the number of parameters. The local data set containing labeled data samples at the $\indexED$th \ac{ED} as $\{(\sampleData[\indexSampleData], \sampleLabel[\indexSampleData] )\}\in\dataset[\indexED]$, where $\sampleData[\indexSampleData]$ and $\sampleLabel[\indexSampleData]$ are the $\indexSampleData$th data sample and its associated label respectively. In this study, unlike to a classical \ac{FEEL} problem, to capture the model test accuracy for each \ac{ED} under heterogeneous data distribution, we define a personalized global loss function at the $\indexED$th \ac{ED} for a given $\modelParametersAtIteration[\indexCommunicationRound][\indexED]$ as
\begin{align}\label{usercentriclossfunction}
    \GlobalusercentricLF[\indexED][{\modelParametersAtIteration[\indexCommunicationRound][\indexED]}]=\frac{1}{|\localdataset[\indexED]|}\sum_{\forall(\sampleData[\indexSampleData], \sampleLabel[\indexSampleData] )\in\localdataset[\indexED]}\lossFunctionSampleLocal[{{\modelParametersAtIteration[\indexCommunicationRound][\indexED]},\sampleData[\indexSampleData],\sampleLabel[\indexSampleData]}][\indexED]~,
 \end{align}
where $\localdataset[\indexED]=\{(\sampleData[\indexSampleData],\sampleLabel[\indexSampleData])\in\completeData|\sampleLabel[\indexSampleData]\in\localClasses[\indexED]\}$ for $\completeData=\dataset[1]\cup\dataset[2]\cup\mydots\cup\dataset[K]$, and $\localClasses[\indexED]$ is the set of distinct labels in the dataset of the $\indexED$th \Ac{ED}. $\lossFunctionSampleLocal[{{\modelParametersAtIteration[\indexCommunicationRound][\indexED]},\sampleData[\indexSampleData],\sampleLabel[\indexSampleData]}][\indexED]$ is the sample loss function that measures the labelling error for $(\sampleData[\indexSampleData], \sampleLabel[\indexSampleData])$ for the parameters ${\modelParametersAtIteration[\indexCommunicationRound][\indexED]}$ at the $\indexED$th \ac{ED}.
The personalized \ac{FL} problem can then be defined as 
% $\modelParametersOptimal=\arg\min_{\modelParameters}  \GlobalusercentricLF[\indexED][{\modelParametersAtIteration[\indexCommunicationRound][\indexED]}]$.
\begin{align}
	\modelParametersOptimal=\arg\min_{\modelParameters}  \GlobalusercentricLF[\indexED][{\modelParametersAtIteration[\indexCommunicationRound][\indexED]}]~.
% 	\frac{1}{|\localdataset[\indexED]|}\sum_{\forall(\sampleData[\indexSampleData],\sampleLabel[\indexSampleData])\in\localdataset[\indexED]}\lossFunctionSampleLocal[{{\modelParameters},\sampleData[\indexSampleData],\sampleLabel[\indexSampleData]}][\indexED]
 	\label{eq-opt}
\end{align}
To solve \eqref{eq-opt}, a full-batch gradient 
descent with the learning rate $\learningRate$ is given by $\modelParametersAtIteration[\indexCommunicationRound+1][\indexED]= \modelParametersAtIteration[\indexCommunicationRound][\indexED] - \learningRate  \localGradient[\indexED][\indexCommunicationRound]$, and
% \begin{align}
%     \modelParametersAtIteration[\indexCommunicationRound+1][\indexED]= \modelParametersAtIteration[\indexCommunicationRound][\indexED] - \learningRate  \localGradient[\indexED][\indexCommunicationRound]~,
% \end{align}
\begin{align}\label{localgradientfunction}
    \localGradient[\indexED][\indexCommunicationRound]=\nabla\GlobalusercentricLF[\indexED][{\modelParametersAtIteration[\indexCommunicationRound][\indexED]}]=\frac{1}{|\localdataset[\indexED]|}\sum_{\forall(\sampleData[\indexSampleData],\sampleLabel[\indexSampleData])\in\localdataset[\indexED]}\nabla\lossFunctionSampleLocal[{\modelParametersAtIteration[\indexCommunicationRound][\indexED],\sampleData[\indexSampleData],\sampleLabel[\indexSampleData]}][\indexED]~,
\end{align}
where the $\indexGradient$th element of $\localGradient[\indexED][\indexCommunicationRound]$ is $\ElementsOfGradient[\indexED,\indexGradient][\indexCommunicationRound]$, which is the gradient of $\GlobalusercentricLF[\indexED][{\modelParametersAtIteration[\indexCommunicationRound][\indexED]}]$ with respect to  $\modelParametersAtIterationEle[\indexCommunicationRound][\indexED,\indexGradient]$.

% \begin{figure}[t]
% 	\centering
% 	\includegraphics[width =3.5in]{txrx.pdf}\label{subfig:txrx}
% 	\caption{FSK-MV \cite{sahinCommnet_2021}.}
% 	\label{fig:mvfsk}
% \end{figure}
In this study, our main goal is to solve \eqref{eq-opt} in a wireless network consisting of multiple cells, where the data sharing among \acp{ED} is not allowed to promote data privacy. To this end, we consider \ac{FEEL} and reduce the communication latency by adopting an \ac{OAC} scheme, i.e., \ac{FSK-MV} \cite{sahinCommnet_2021}, which is originally proposed  in the \ac{UL} for a single cell (i.e., $\numberOfEdgeServers=1$).  With this scheme, the $\indexED$th \ac{ED} first calculates the local stochastic gradient as
\begin{align}\label{statisticgradient}
    \localGradientStat[\indexED][\indexCommunicationRound]=\frac{1}{\batchSize}\sum_{\forall(\sampleData[\indexSampleData],\sampleLabel[\indexSampleData])\in\datasetBatch[\indexED]}\nabla\lossFunctionSampleLocal[{\modelParametersAtIteration[\indexCommunicationRound][\indexED],\sampleData[\indexSampleData],\sampleLabel[\indexSampleData]}][\indexED]~,
\end{align}
where  $\localGradientStat[\indexED][\indexCommunicationRound]$ is the local gradient where its $\indexGradient$th element is $\localGradientStatElement[\indexED,\indexGradient][\indexCommunicationRound]$ and $\datasetBatch[\indexED]\subset\dataset[\indexED]$ is the selected data batch from the local data set with the batch size, $\batchSize=|\datasetBatch[\indexED]|$. Each \ac{ED} then obtains the transmit symbols in the \ac{UL} as follows: Consider a mapping from $\indexGradient\in\lbrace 1,...,\numberOfModelParametersED[]\rbrace$ to the distinct pairs $(\voteInFrequency[+],\voteInTime[+])$ and $(\voteInFrequency[-],\voteInTime[-])$ for $\voteInFrequency[+],\voteInFrequency[-]\in\{0,1,\mydots,\numberOfActiveSubcarriers-1\}$ and $\voteInTime[+],\voteInTime[-]\in\{0,1,\mydots,\numberOfOFDMSymbols-1\}$. Based on the value of $\localGradientSignElement[\indexED,\indexGradient][\indexCommunicationRound]\triangleq\sign{(\localGradientStatElement[\indexED,\indexGradient][\indexCommunicationRound])}$, the $\indexED$th \ac{ED} calculates the symbol $\transmittedSymbolAtSubcarrierED[{\indexED,\voteInFrequency[+],\voteInTime[+]}]$ and $\transmittedSymbolAtSubcarrierED[{\indexED,\voteInFrequency[-],\voteInTime[-]}]$, $\forall \indexGradient$, as
\begin{align}
\transmittedSymbolAtSubcarrierED[{\indexED,\voteInFrequency[+],\voteInTime[+]}]=\sqrt{\symbolEnergy}\randomizationSymbolAtSubcarrierED[{\indexED,\indexGradient}]\indicatorFunction[{\localGradientSignElement[\indexED,\indexGradient][\indexCommunicationRound] =1}]~,
\label{eq:symbolPlusED}
\end{align}
and
\begin{align}
\transmittedSymbolAtSubcarrierED[{\indexED,\voteInFrequency[-],\voteInTime[-]}]=\sqrt{\symbolEnergy}\randomizationSymbolAtSubcarrierED[{\indexED,\indexGradient}]\indicatorFunction[{\localGradientSignElement[\indexED,\indexGradient][\indexCommunicationRound] =-1}]~,
\label{eq:symbolMinusED}
\end{align}
respectively, where $\randomizationSymbolAtSubcarrierED[{\indexED,\indexGradient}]$ is a random \ac{QPSK} symbol and $\symbolEnergy=2$ is the symbol energy.  Note that a long-term power constraint, used for \ac{OBDA} \cite[Eq. 9 and Eq. 10]{Guangxu_2021}, is not needed for \ac{FSK-MV} as the \ac{OFDM} symbol energy does not change  as a function of \ac{CSI}  with \ac{FSK-MV}.
%It is assumed that the multi-path channels between the \Acp{ES} and \Acp{ED} are independent.
%transmit the signs of their local stochastic gradients, $\localGradientSign[\indexED][\indexCommunicationRound]$, to the server instead of the actual values of $\localGradient[\indexED][\indexCommunicationRound]$ for $\indexED=1,\mydots,\numberOfEdgeDevices$. The $\indexGradient$th element of $\localGradientSign[\indexED][\indexCommunicationRound]$ is $\localGradientSignElement[\indexED,\indexGradient][\indexCommunicationRound]=\sign{(\ElementsOfGradient[\indexED,\indexGradient][\indexCommunicationRound])}$. The \Ac{MV} of the $\indexED$th \Ac{ED} for the $\indexGradient$th gradient is {\color{blue}
% \begin{align}
    % \signofpsiElement[\indexCommunicationRound][{\indexED,\indexGradient}]=\sign(\sum^{\numberofconnectedED}_{\indexED=1}\localGradientSignElement[\indexED,\indexGradient][\indexCommunicationRound])~,\qquad \forall \indexGradient \in \lbrace 1,...,q\rbrace~.
% \end{align}
The \ac{ES} receives the superposed symbols for a given $\indexGradient$, respectively, as follows:
\begin{align}
    \ESreceivedsignal[\indexCommunicationRound][{\indexES,\voteInFrequency[+],\voteInTime[+]}]=\sum_{\substack{\indexED=1,\localGradientStatElement[\indexED,\indexGradient][\indexCommunicationRound]>0}} ^{\numberOfEdgeDevices}\sqrt{\receivedPowerOfED[\indexES][{\indexED}]}\ULchannel[\indexCommunicationRound][{\indexES,\indexED,\voteInFrequency[+],\voteInTime[+]}]\transmittedSymbolAtSubcarrierED[{\indexED,\voteInFrequency[+],\voteInTime[+]}]+\noiseESplus[\indexCommunicationRound][{\indexES,\voteInFrequency[+],\voteInTime[+]}]~,\nonumber
\end{align}
\begin{align}
 \ESreceivedsignal[\indexCommunicationRound][{\indexES,\voteInFrequency[-],\voteInTime[-]}]=\sum_{\substack{\indexED=1,\localGradientStatElement[\indexED,\indexGradient][\indexCommunicationRound]<0}} ^{\numberOfEdgeDevices}\sqrt{\receivedPowerOfED[\indexES][{\indexED}]}\ULchannel[\indexCommunicationRound][{\indexES,\indexED,\voteInFrequency[-],\voteInTime[-]}]\transmittedSymbolAtSubcarrierED[{\indexED,\voteInFrequency[-],\voteInTime[-]}]+\noiseESplus[\indexCommunicationRound][{\indexES,\voteInFrequency[-],\voteInTime[-]}]~.\nonumber
\end{align}
The superposed symbols at the \ac{ES}  are then compared with an energy detector for the $\indexGradient$th gradient to detect the \ac{MV} as $\signofdeltaElement[\indexCommunicationRound][{\indexES,\indexGradient}]=\sign(\deltaofr[\indexCommunicationRound][{\indexES,\indexGradient}])$ for $\forall \indexGradient$, where
%\begin{align}\label{eq:detector}
    %\signofdeltaElement[\indexCommunicationRound][{\indexES,\indexGradient}]=\sign(\deltaofr[\indexCommunicationRound][{\indexES,\indexGradient}])~,\qquad \forall \indexGradient \in \lbrace 1,...,\numberOfModelParametersED[]\rbrace~,
%\end{align}
$\deltaofr[\indexCommunicationRound][{\indexES,\indexGradient}]\triangleq|{\ESreceivedsignal[\indexCommunicationRound][{\indexES,\voteInFrequency[+],\voteInTime[+]}]}|^2-|{\ESreceivedsignal[\indexCommunicationRound][{\indexES,\voteInFrequency[-],\voteInTime[-]}]}|^2$. Finally, the \ac{ES} transmits the \acp{MV}, i.e., $\signofdelta[\indexCommunicationRound][{\indexES}]=[\signofdeltaElement[\indexCommunicationRound][\indexES,1],\mydots,\signofdeltaElement[\indexCommunicationRound][{\indexES,\numberOfModelParametersED[]}]]^{\rm T}$, to the \acp{ED} and the model parameters at the $\indexED$th \ac{ED} are updated as
\begin{align} 
  {\modelParametersAtIteration[\indexCommunicationRound+1][\indexED]} = \modelParametersAtIteration[\indexCommunicationRound][\indexED] - \learningRate  \signofpsi[\indexCommunicationRound][{\indexED}]~.
  \label{eq:updateRuleForMC}
\end{align}
This procedure is repeated for $\communicationRounds$ communication rounds. 
% The corresponding transmitter and receiver block diagrams for \ac{UL} \ac{OAC} with \ac{FSK-MV} are given in \figurename~\ref{fig:mvfsk}. %With the original \ac{FSK-MV}, \ac{OAC} is proposed only in the \ac{UL}, the \ac{DL} communication is assumed to be ideal, and analysis is done for a single cell. 

%\subsection{Performance Metrics}
\begin{figure}[t]
	\centering
	\subfloat[UL OAC.]{\includegraphics[width =2.25in]{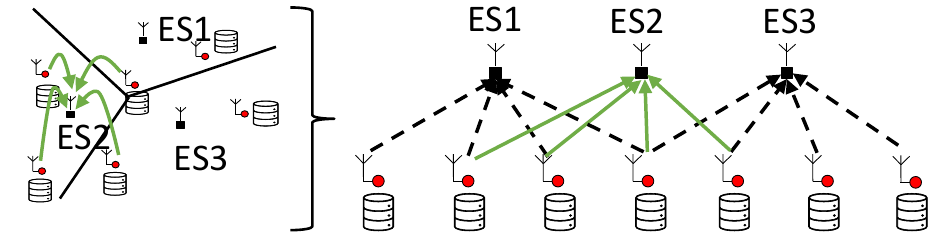}\label{subfig:ul_oac}}\\
	\subfloat[DL OAC.]{\includegraphics[width =2.25in]{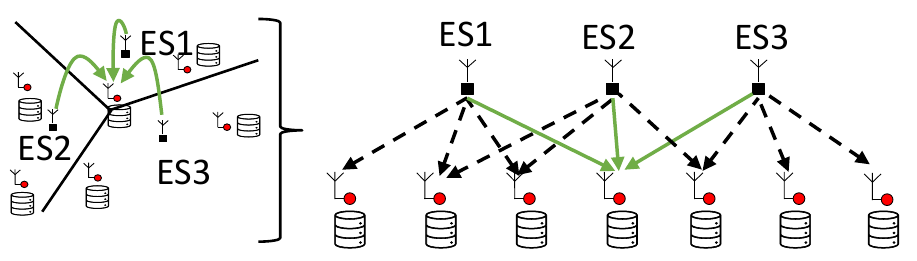}\label{subfig:dl_oac}}	
	\caption{The interference across the cells (i.e., among the EDs or the ESs) is exploited in both UL and DL for the gradient aggregation.}
	\label{fig:mcaircomp}
\end{figure}
\section{Multi-cell Over-the-Air Computation}
\label{sec:mcAirComp}
One of the major advantages of \ac{FSK-MV} over other state-of-the-art \ac{OAC} schemes (e.g., \ac{OBDA} \cite{Guangxu_2021}) is that \acp{ED} and \acp{ES} do not need to utilize the \ac{CSI}. Also, it does not require precise time-synchronization among the transmitters since the computation with \ac{FSK-MV} is achieved through a non-coherent detection in the frequency domain. 
These unique features enable us to extend \ac{FSK-MV} in a multi-cell environment as the interference  in both \ac{UL} and \ac{DL} can be exploited for computations: In the \ac{UL}, the transmitted symbols from an \ac{ED} superpose not only with the other \acp{ED} in the cell, but also with the ones at the neighboring cells. Therefore,  the \ac{MV} calculation at the \acp{ES} can exploit the interference from the \acp{ED} located at the neighboring cells as illustrated in \figurename~\ref{fig:mcaircomp}\subref{subfig:ul_oac}. Similarly, in the \ac{DL}, an \ac{ED} (e.g., a cell-edge \ac{ED}) can receive signals from multiple \acp{ES}. Hence, the inter-cell interference in the \ac{DL} can also be used for the \ac{MV} calculation at the \acp{ED} as depicted in \figurename~\ref{fig:mcaircomp}\subref{subfig:dl_oac}. We discuss the operations at \acp{ED} and \acp{ES} in the following subsections in detail.  The proposed framework is also outlined in Algorithm~\ref{alg:mcAirComp}.

\begin{algorithm}[t]
	\scriptsize
	\caption{Multi-cell over-the-air computation}\label{alg:mcAirComp}
    \SetKwInput{KwInput}{Input}                % Set the Input
    \SetKwInput{KwOutput}{Output}              % set the Output
    \SetKwComment{Comment}{/* }{}
    
	\DontPrintSemicolon
	
	%\KwInput{$\learningRate$, $\communicationRounds$}
%	\KwOutput{$\{\modelParametersAtIteration[\communicationRounds][\indexED],\forall\indexED\}$}
	
	% Set Function Names
	\SetKwFunction{FMain}{}
	\SetKwProg{Fn}{Function}{ multiCellOAC}{}
	\Fn{\FMain}{
 \For{$\indexCommunicationRound=1:\communicationRounds$} {
        \Comment*[l]{Processing @ EDs}
            \For{$\indexED=1:\numberOfEdgeDevices$}{
    	    Determine $\EDreceivedsignal[\indexCommunicationRound][{\indexED,\voteInFrequency[+],\voteInTime[+]}],\EDreceivedsignal[\indexCommunicationRound][{\indexED,\voteInFrequency[-],\voteInTime[-]}]$, $\forall\indexGradient$\;
    	    Detect the \ac{MV} at the \ac{ED}, i.e., $\signofpsiElement[\indexCommunicationRound][{\indexED,\indexGradient}]$, $\forall\indexGradient$\;
           Update the model parameter
           $  {\modelParametersAtIteration[\indexCommunicationRound+1][\indexED]} = \modelParametersAtIteration[\indexCommunicationRound][\indexED] - \learningRate  \signofpsi[\indexCommunicationRound][{\indexED}]~.$\;            
     	    Calculate local gradients based on \eqref{statisticgradient}\;
     	    Calculate $\transmittedSymbolAtSubcarrierED[{\indexED,\voteInFrequency[+],\voteInTime[+]}],\transmittedSymbolAtSubcarrierED[{\indexED,\voteInFrequency[-],\voteInTime[-]}]$ , $\forall\indexGradient$\;
     	    %Calculate the OFDM symbols\;
     	    }
     	    \Comment*[l]{Aggregation in the uplink channel}
     	    EDs transmit the corresponding \ac{OFDM} symbols simultaneously\;
     	    ESs receive the superposed \ac{OFDM} symbols in the uplink\;
     	    \Comment*[l]{Processing @ ESs}
            \For{$\indexES=1:\numberOfEdgeServers$}{
            Determine $\ESreceivedsignal[\indexCommunicationRound][{\indexES,\voteInFrequency[+],\voteInTime[+]}],\ESreceivedsignal[\indexCommunicationRound][{\indexES,\voteInFrequency[-],\voteInTime[-]}]$, $\forall\indexGradient$\;
    	    Detect the MV at the \ac{ES}, i.e., $\signofdeltaElement[\indexCommunicationRound][{\indexES,\indexGradient}]$, $\forall\indexGradient$ \;
    	    Calculate $\transmittedSymbolAtSubcarrierES[{\indexES,\voteInFrequency[+],\voteInTime[+]}],\transmittedSymbolAtSubcarrierES[{\indexES,\voteInFrequency[-],\voteInTime[]}]$, $\forall\indexGradient$\;   
    	    %Calculate the OFDM symbols\;
    	    }
            \Comment*[l]{Aggregation in the downlink channel}
     	    ESs transmit the corresponding \ac{OFDM} symbols simultaneously\;
     	    EDs receive the superposed \ac{OFDM} symbols in the downlink\;    
            }
         }
\end{algorithm}

\subsection{Uplink OAC with FSK-MV}
In the \ac{UL}, the expressions given for the transmitted symbols from the \acp{ED} and the superposed symbols at the \ac{ES} with \ac{FSK-MV}  for a single cell, discussed in Section~\ref{subsec:learning},  also hold in a multi-cell environment for $\numberOfEdgeServers>1$. After the $\indexES$th \ac{ES} calculates the vector $\signofdelta[\indexCommunicationRound][\indexES]$, $\forall\indexES$, the \ac{DL} \ac{OAC} starts.

\subsection{Downlink OAC with FSK-MV}

\subsubsection{Edge Servers - Transmitter}
Similar to the \ac{UL} \ac{OAC}, we first consider a distinct pairs $(\voteInFrequency[+],\voteInTime[+])$ and $(\voteInFrequency[-],\voteInTime[-])$ corresponding to the $\indexGradient$th gradient. Based on the value of $\signofdelta[\indexCommunicationRound][{\indexES}]$, at the $\indexCommunicationRound$th communication round, the $\indexES$th \Ac{ES} calculates the symbol $\transmittedSymbolAtSubcarrierED[{\indexES,\voteInFrequency[+],\voteInTime[+]}]$ and $\transmittedSymbolAtSubcarrierED[{\indexES,\voteInFrequency[-],\voteInTime[-]}]$, $\forall \indexGradient$, as
\begin{align}
\transmittedSymbolAtSubcarrierES[{\indexES,\voteInFrequency[+],\voteInTime[+]}]=\sqrt{\symbolEnergy}\randomizationSymbolAtSubcarrierES[{\indexES,\indexGradient}]\indicatorFunction[{\signofdeltaElement[\indexCommunicationRound][{\indexES,\indexGradient}] =1}]~,
\label{eq:symbolPlusED1}
\end{align}
and
\begin{align}
\transmittedSymbolAtSubcarrierES[{\indexES,\voteInFrequency[-],\voteInTime[-]}]=\sqrt{\symbolEnergy}\randomizationSymbolAtSubcarrierES[{\indexES,\indexGradient}]\indicatorFunction[{\signofdeltaElement[\indexCommunicationRound][{\indexES,\indexGradient}]=-1}]~,
\label{eq:symbolMinusED1}
\end{align}
respectively, where $\randomizationSymbolAtSubcarrierES[{\indexES,\indexGradient}]$ is a random \ac{QPSK} symbol. All \acp{ES} calculate the corresponding \ac{OFDM} symbols and transmit them simultaneously for \ac{DL} \ac{OAC}.
\subsubsection{Edge Device - Receiver}
%Let $\numberOFEDsForOptionOne$ and $\numberOFEDsForOptionSecond=\numberofconnectedED-\numberOFEDsForOptionOne$ be the number of \Acp{ED} with $1$ and $-1$ votes, respectively. Also, 
In the \ac{DL}, the superposed symbols  at the  $\indexED$th \ac{ED} for all $\indexGradient$ can be expressed as
\begin{align}
   \EDreceivedsignal[\indexCommunicationRound][{\indexED,\voteInFrequency[+],\voteInTime[+]}]=\sum_{\indexES=1,\deltaofr[\indexCommunicationRound][\indexES,\indexGradient]>0} ^{\numberOfEdgeServers}\sqrt{\receivedPowerOfES[\indexES][\indexED]}\DLchannel[\indexCommunicationRound][{\indexES,\indexED,\voteInFrequency[+],\voteInTime[+]}]\transmittedSymbolAtSubcarrierES[{\indexES,\voteInFrequency[+],\voteInTime[+]}]+\noiseEDplus[\indexCommunicationRound][{\indexES,\voteInFrequency[+],\voteInTime[+]}]~,\nonumber
\end{align}
and
\begin{align}
    \EDreceivedsignal[\indexCommunicationRound][{\indexED,\voteInFrequency[-],\voteInTime[-]}]=\sum_{\indexES=1,\deltaofr[\indexCommunicationRound][\indexES,\indexGradient]<0} ^{\numberOfEdgeServers}\sqrt{\receivedPowerOfES[\indexES][\indexED]}\DLchannel[\indexCommunicationRound][{\indexES,\indexED,\voteInFrequency[-],\voteInTime[-]}]\transmittedSymbolAtSubcarrierES[{\indexES,\voteInFrequency[-],\voteInTime[-]}]+\noiseEDplus[\indexCommunicationRound][{\indexES,\voteInFrequency[-],\voteInTime[-]}]~.\nonumber
\end{align}
The energy detector at the $\indexED$th \ac{ED} then detects the \ac{MV}  for the $\indexGradient$th gradient as $\signofpsiElement[\indexCommunicationRound][{\indexED,\indexGradient}]=\sign(\psiforp[\indexCommunicationRound][{\indexED,\indexGradient}])$ for $\forall \indexGradient$, where
%\begin{align} \label{EDconvergence}
  %  \signofpsiElement[\indexCommunicationRound][{\indexED,\indexGradient}]=\sign(\psiforp[\indexCommunicationRound][{\indexED,\indexGradient}])~,\qquad \forall \indexGradient \in \lbrace 1,...,\numberOfModelParametersED[]\rbrace~,
%\end{align}
$\psiforp[\indexCommunicationRound][{\indexED,\indexGradient}]\triangleq|{\EDreceivedsignal[\indexCommunicationRound][{\indexED,\voteInFrequency[+],\voteInTime[+]}]}|^2-|{\EDreceivedsignal[\indexCommunicationRound][{\indexED,\voteInFrequency[-],\voteInTime[-]}]}|^2$. Subsequently, the $\indexED$th \ac{ED} calculates the \ac{MV} vector, i.e., $\signofpsi[\indexCommunicationRound][{\indexED}]=[\signofpsiElement[\indexCommunicationRound][{\indexED,1}],\mydots,\signofpsiElement[\indexCommunicationRound][{\indexED,\numberOfModelParametersED[]}]]^{\rm T}$ and updates its parameters as in \eqref{eq:updateRuleForMC}. Hence, the parameters at the \acp{ED} are updated based on the received signals from multiple \acp{ES}.

\subsection{Convergence Analysis}
For the convergence analysis, we consider several standard assumptions made in the literature \cite{Bernstein_2018, Guangxu_2021}:
\begin{assumption}[Bounded loss function] The loss function has a lower bound for some value $\lossFunctionGlobalMinimum$, i.e.,
	\rm 
	$\GlobalusercentricLF[\indexED][\modelParameters]\ge \lossFunctionGlobalMinimum$, $\forall\modelParameters$ \cite{Bernstein_2018}. 
\end{assumption}
\begin{assumption}[Smoothness] 
	\rm 
	Let $\localGradientNoIndex[\indexED]$ be the gradient of the personalized global loss function $\GlobalusercentricLF[\indexED][\modelParameters]$ evaluated at $\modelParameters$. For all $\modelParameters$ and $\modelParameters'$, the expression given by
	\begin{align}
		\left| \GlobalusercentricLF[\indexED][{\modelParameters'}] - (\lossFunctionGlobal[\modelParameters]-\localGradientNoIndex[\indexED]^{\rm T}(\modelParameters'-\modelParameters)) \right| \le \frac{1}{2}\sum_{\indexGradient=1}^{\numberOfModelParametersED[]} \nonnegativeConstantsEle[\indexGradient](\modelParametersEle[\indexGradient]'-\modelParametersEle[\indexGradient])^2~,
		\nonumber
	\end{align}	
	holds for a non-negative constant vector
	$\nonnegativeConstants=[\nonnegativeConstantsEle[1],\mydots,\nonnegativeConstantsEle[{\numberOfModelParametersED[]}]]^{\rm T}$.
\end{assumption}
\begin{assumption}[Variance bound] \rm Assume that the estimated gradient is an unbiased estimate of the true gradient, $\expectationOperator[{\localGradientStatNoIndex[\indexED]}][]=\localGradientNoIndex[\indexED],~\forall\indexED$, and the variance of each component of them is bounded as $\expectationOperator[{(\localGradientStatNoIndexElement[\indexED,\indexGradient]-\globalGradientElementNoIndex[\indexED,\indexGradient])^2}][]\le\varianceBoundEle[\indexGradient]^2/\batchSize,~\forall\indexED,\indexGradient$,
	%\rm The stochastic  gradient estimates $\localGradientNoIndex[\indexED]=[\localGradientNoIndexElement[\indexED,1],\mydots,\localGradientNoIndexElement[\indexED,{\numberOfModelParametersED[]}]]^{\rm T}=\nabla \lossFunctionLocal[\indexED][{\modelParametersAtIteration[\indexCommunicationRound][]}]$, $\forall\indexED$, are independent and  unbiased estimates of $\globalGradientNoIndex=[\globalGradientElementNoIndex[1],\mydots,\globalGradientElementNoIndex[{\numberOfModelParametersED[]}]]^{\rm T}=\nabla\lossFunctionGlobal[{\modelParameters}]$ with a coordinate bounded variance, i.e.,
% \begin{align}
%  		\expectationOperator[{\localGradientStatNoIndex[\indexED]}][]&=\localGradientNoIndex[\indexED],~\forall\indexED,\\	\expectationOperator[{(\localGradientStatNoIndexElement[\indexED,\indexGradient]-\globalGradientElementNoIndex[\indexED,\indexGradient])^2}][]&\le\varianceBoundEle[\indexGradient]^2/\batchSize,~\forall\indexED,\indexGradient,
%   	\end{align}
where $\varianceBound = [\varianceBoundEle[1],\mydots,\varianceBoundEle[{\numberOfModelParametersED[]}]]^{\rm T}$ is a non-negative constant  vector.
 \end{assumption}
 \begin{assumption}[Unimodal, symmetric gradient noise]
 	\rm
 	For any given $\modelParameters$, the elements of the vector $\localGradientNoIndex[\indexED]$, $\forall\indexED$, has an unimodal distribution that is also symmetric around its mean.
 \end{assumption}

We also assume that  the number of \acp{ED} that are connected to an \ac{ES} and the number of \acp{ES} that are connected to an \ac{ED} are fixed and denoted as  $\numberofconnectedED\le\numberOfEdgeDevices$ and  $\numberofconnectedES\le \numberOfEdgeServers$, respectively (i.e., fixed-connectivity assumption). This assumption is due to the large-scale fading in wireless channels, e.g., an \ac{ES} can receive the strong signals from the \acp{ED} located at its adjacent \acp{ES}, but the ones from far cells are likely to be attenuated due to the large link distance. Based on this assumption,
let $\mathcal{K}_{\indexES}$ be the set of all \Acp{ED} that are connected to the $\indexES$th \Ac{ES} and $\mathcal{S}_{\indexED}$ be the set of all \Acp{ES} that are connected to the $\indexED$th \Ac{ED}, where $|\mathcal{K}_\indexES|=\numberofconnectedED$, $\forall\indexED$, and $|\mathcal{S}_\indexED|=\numberofconnectedES$, $\forall\indexES$. We set the received power $\receivedPowerOfED[\indexES][{\indexED}]=1$ for $\indexED\in{\mathcal{K}_\indexES}, \forall\indexES$, otherwise $0$, and $\receivedPowerOfES[\indexES][{\indexED}]=1$ for $\indexES\in{\mathcal{S}_{\indexED}}, \forall\indexED$, otherwise $0$. This assumption does not hold for an irregular deployment. Nevertheless, it leads us to provide insight into multi-cell \ac{OAC} with  a tractable analysis since it results in  $|{\ESreceivedsignal[\indexCommunicationRound][{\indexES,\voteInFrequency[+],\voteInTime[+]}]}|^2$ and $|{\ESreceivedsignal[\indexCommunicationRound][{\indexES,\voteInFrequency[-],\voteInTime[-]}]}|$ to be exponential random variables with the means $\meanOptionOneES=\symbolEnergy\numberOFEDsForOptionOne+\noiseVariance$ and $\meanOptionTwoES=\symbolEnergy\constantPowerOfED\numberOFEDsForOptionSecond+\noiseVariance$, respectively, 
%where $\numberOFEDsForOptionOne$ is the number of \acp{ED} with $+1$ vote and $\numberOFEDsForOptionSecond\triangleq\numberofconnectedED-\numberOFEDsForOptionOne$ is the number of \acp{ED} with $-1$ vote.
where $\numberOFEDsForOptionOne$ and $\numberOFEDsForOptionSecond$ are the cardinalities of the sets $\{{\localGradientSignElement[\indexED,\indexGradient][\indexCommunicationRound] =+1}|\indexED\in\mathcal{K}_{\indexES}\}$ and $\{{\localGradientSignElement[\indexED,\indexGradient][\indexCommunicationRound] =-1}|\indexED\in\mathcal{K}_{\indexES}\}$, respectively.
Also, $|{\EDreceivedsignal[\indexCommunicationRound][{\indexED,\voteInFrequency[+],\voteInTime[+]}]}|^2$ and $|{\EDreceivedsignal[\indexCommunicationRound][{\indexED,\voteInFrequency[-],\voteInTime[-]}]}|^2$ become exponential random variables  with the means $\meanOptionOne=\symbolEnergy\constantPowerOfES\numberOFESsForOptionOne+\noiseVarianced$ and $\meanOptionTwo=\symbolEnergy\constantPowerOfES\numberOFESsForOptionSecond+\noiseVarianced$, respectively, where $\numberOFESsForOptionOne$ and $\numberOFESsForOptionSecond$ are the cardinalities of the sets $\{{\signofdeltaElement[\indexCommunicationRound][{\indexES,\indexGradient}]=+1}|\indexES\in\mathcal{S}_{\indexED}\}$ and $\{{\signofdeltaElement[\indexCommunicationRound][{\indexES,\indexGradient}]=-1}|\indexES\in\mathcal{S}_{\indexED}\}$, respectively.
%the numbers of \acp{ES} with $+1$ and $-1$ votes are $\numberOFESsForOptionOne$ and $\numberOFESsForOptionSecond\triangleq\numberofconnectedES-\numberOFESsForOptionOne$, respectively. 
The distributions of $\deltaofr[\indexCommunicationRound][\indexES,\indexGradient]$ and $\psiforp[\indexCommunicationRound][\indexED,\indexGradient]$ can then be calculated as $\deltaofr[\indexCommunicationRound][\indexES,\indexGradient]\sim f(x,\meanOptionOneES,\meanOptionTwoES)$ and $\psiforp[\indexCommunicationRound][\indexED,\indexGradient]	\sim f(y,\meanOptionOne,\meanOptionTwo)$ respectively, where
$f(x,\mu_1,\mu_2)$ is  ${\constante^{-\frac{x}{\mu_1}}}/{(\mu_1+\mu_2)}$ for $ x> 0$, and otherwise it is ${\constante^{-\frac{x}{\mu_2}}}/{(\mu_1+\mu_2)}$ \cite{sahinCommnet_2021}. 
% 	 f(x,\mu_1,\mu_2)\triangleq \begin{cases} 
% 			\frac{\constante^{-\frac{x}{\mu_1}}}{\mu_1+\mu_2}, & x> 0 \\
% 			\frac{\constante^{-\frac{x}{\mu_2}}}{\mu_1+\mu_2}, & x\le0 
% 		\end{cases}~.
% \end{align}

\begin{theorem}
\label{th:main}
	\rm For $\learningRate=\frac{1}{\communicationRounds}$ and $\batchSize=\frac{\communicationRounds}{\gamma}$, the convergence rate of multi-cell \ac{OAC} with \ac{FSK-MV} in Rayleigh fading channel is
	\begin{align}
    &\expectationOperator[\frac{1}{\communicationRounds}{\sum^{\communicationRounds-1}_{\indexCommunicationRound=0}\norm{\localGradient[\indexED][\indexCommunicationRound]}_1}][]\nonumber\\&\leq\frac{1}{(\numberOfEdgeDevices-2A){\sqrt{\communicationRounds}}}\left(\lossFunctionLocal[\indexED][{\modelParametersAtIteration[0][\indexED]}]-\lossFunctionGlobalMinimum+\frac{1}{2}\numberOfEdgeDevices\norm{\constantvector}_1+2\sqrt{\gamma}B\frac{\sqrt{2}}{3}\norm{\varianceBound}_1\right),\label{eq:convergence}
\end{align}
where $\gamma$ is a positive integer, $A$ and $B$ are defined as
$A\triangleq\frac{1}{1+\noiseVarianced}-B$
%$A=\frac{\symbolEnergy^2\constantPowerOfED\numberofconnectedES+\noiseVarianced}{\symbolEnergy^2\constantPowerOfED\numberofconnectedES+2\noiseVarianced}-B$,
and
$B\triangleq\frac{\constantPowerOfED\numberofconnectedES(\noiseVariance+\symbolEnergy\constantPowerOfED\numberofconnectedED)}{\symbolEnergy(\constantPowerOfED\numberofconnectedES+2\noiseVarianced)(\constantPowerOfED\numberofconnectedED+2\noiseVariance)}$, respectively.
% $B=\frac{\symbolEnergy\constantPowerOfED\numberofconnectedES\noiseVariance+\symbolEnergy\constantPowerOfED^2\numberofconnectedES\numberofconnectedED}{(\symbolEnergy\constantPowerOfED\numberofconnectedES+2\noiseVarianced)(\symbolEnergy\constantPowerOfED\numberofconnectedED+2\noiseVariance)}$.
\end{theorem}

\begin{IEEEproof}
The proof relies on the strategy used in \cite{Bernstein_2018}. % which relates the norm of the global loss function to the expected improvement made in each communication round. 
By using Assumption~2 and using \eqref{eq:updateRuleForMC}, it can be shown that
% \begin{align}
%     &\lossFunctionLocal[\indexED][{\modelParametersAtIteration[\indexCommunicationRound+1][\indexED]}]-\lossFunctionLocal[\indexED][{\modelParametersAtIteration[\indexCommunicationRound][\indexED]}]\nonumber\\&\leq{\localGradient[\indexED][\indexCommunicationRound]}^T(\modelParametersAtIteration[\indexCommunicationRound+1][\indexED]-\modelParametersAtIteration[\indexCommunicationRound][\indexED])+\frac{1}{2}\sum^{\numberOfEdgeDevices}_{\indexED=1}\sum_{\indexGradient=1}^{\numberOfModelParametersED[]}{\indexofconstantvector[\indexGradient]({\modelParametersAtIteration[\indexCommunicationRound][\indexED]}}-\modelParametersAtIteration[\indexCommunicationRound][\indexED])\nonumber\\&=-\learningRate\localGradient[\indexED][\indexCommunicationRound]\signofpsi[\indexCommunicationRound][\indexED]+\frac{{\learningRate}^2}{2}\sum^{\numberOfEdgeDevices}_{\indexED=1}\sum_{\indexGradient=1}^{\numberOfModelParametersED[]}{\indexofconstantvector[\indexGradient]}=\learningRate\numberOfEdgeDevices\norm{\localGradient[\indexED][\indexCommunicationRound]}_1+\frac{{\learningRate}^2}{2}\numberOfEdgeDevices\norm{\constantvector}_1\nonumber\\\nonumber&+2\learningRate\sum^{\numberOfEdgeDevices}_{\indexED=1}\sum_{\indexGradient=1}^{\numberOfModelParametersED[]}|\ElementsOfGradient[\indexED,\indexGradient][\indexCommunicationRound]|\indicatorFunction[{\signofpsiElement[\indexCommunicationRound][{\indexED,\indexGradient}]\neq\SignofUsercentricElement[\indexED,\indexGradient][\indexCommunicationRound]}].\nonumber
% \end{align}
%Therefore,
\begin{align}
    &\expectationOperator[{\lossFunctionLocal[\indexED][{\modelParametersAtIteration[\indexCommunicationRound+1][\indexED]}]-\lossFunctionLocal[\indexED][{\modelParametersAtIteration[\indexCommunicationRound][\indexED]}]}][]\leq-\learningRate\numberOfEdgeDevices\norm{\localGradient[\indexED][\indexCommunicationRound]}_1+\frac{{\learningRate}^2}{2}\numberOfEdgeDevices\norm{\constantvector}_1\nonumber\\&~~~~~~~~~~~~~~+2\learningRate\sum^{\numberOfEdgeDevices}_{\indexED=1}\sum_{\indexGradient=1}^{\numberOfModelParametersED[]}|\ElementsOfGradient[\indexED,\indexGradient][\indexCommunicationRound]|\probability[{\signofpsiElement[\indexCommunicationRound][{\indexED,\indexGradient}]\neq\SignofUsercentricElement[\indexED,\indexGradient][\indexCommunicationRound]}]~,\nonumber
\end{align}
 where $\sum^{\numberOfEdgeDevices}_{\indexED=1}\sum_{\indexGradient=1}^{\numberOfModelParametersED[]}|\ElementsOfGradient[\indexED,\indexGradient][\indexCommunicationRound]|\probability[{\signofpsiElement[\indexCommunicationRound][{\indexED,\indexGradient}]\neq\SignofUsercentricElement[\indexED,\indexGradient][\indexCommunicationRound]}]$ is the stochasticity-induced error. Let $\SignofUsercentricElement[\indexED,\indexGradient][\indexCommunicationRound]\triangleq\sign{(\ElementsOfGradient[\indexED,\indexGradient][\indexCommunicationRound])}$ denote the correct decision.
 %and assume that $\SignofUsercentricElement[\indexED,\indexGradient][\indexCommunicationRound]=1$. 
 Also, let $\numberOfESsWithCorrectChoice$ and  $\numberOfEDsWithCorrectChoice$ be binomial random variables for the number of \acp{ES} and the number of \acp{ED} with the correct decision, i.e., 
 $\numberOfESsWithCorrectChoice\sim\binomialRandomVariable[\numberofconnectedES][\correctDecisionED]$ and $\numberOfEDsWithCorrectChoice\sim\binomialRandomVariable[\numberofconnectedED][{\correctDecision[\indexGradient]}]$, where $\correctDecisionED$ and 
$\correctDecision[\indexGradient]$ denote the success probabilities \cite{Bernstein_2018}. The probability $\probabilityIncorrect[\indexED,\indexGradient]\triangleq\probability[{\signofpsiElement[\indexCommunicationRound][{\indexED,\indexGradient}]\neq\SignofUsercentricElement[\indexED,\indexGradient][\indexCommunicationRound]}]$ and the success probability $\correctDecisionED$  can then be written as
\begin{align}
    &\probabilityIncorrect[\indexED,\indexGradient]=\sum^{\numberofconnectedES}_{\numberOFESsForOptionOne=1}\probability[{\signofpsiElement[\indexCommunicationRound][{\indexED,\indexGradient}]\neq\SignofUsercentricElement[\indexED,\indexGradient][\indexCommunicationRound]|\numberOfESsWithCorrectChoice=\numberOFESsForOptionOne}]
   \probability[{\numberOfESsWithCorrectChoice=\numberOFESsForOptionOne}]~,\label{errorP}
\end{align}
and
\begin{align}
    \correctDecisionED=&\sum^{\numberofconnectedED}_{\numberOFEDsForOptionOne=1}\probability[{\signofdeltaElement[\indexCommunicationRound][{\indexES,\indexGradient}]=\SignofUsercentricElement[\indexED,\indexGradient][\indexCommunicationRound]|\numberOfEDsWithCorrectChoice=\numberOFEDsForOptionOne}]\probability[\numberOfEDsWithCorrectChoice=\numberOFEDsForOptionOne]~,\label{EScorrectdecision}%\nonumber
\end{align} 
respectively. Based on the distributions of $\deltaofr[\indexCommunicationRound][\indexES,\indexGradient]$ and $\psiforp[\indexCommunicationRound][\indexED,\indexGradient]$, we calculate the conditional probabilities in \eqref{errorP} and \eqref{EScorrectdecision} as
\begin{align}
    \probability[{\signofpsiElement[\indexCommunicationRound][{\indexED,\indexGradient}]\neq\SignofUsercentricElement[\indexED,\indexGradient][\indexCommunicationRound]|\numberOfESsWithCorrectChoice=\numberOFESsForOptionOne}]&=\frac{\meanOptionTwo}{\meanOptionOne+\meanOptionTwo}~,
    %\nonumber\\&\hspace{-14mm}=\frac{\symbolEnergy\constantPowerOfES\numberOFESsForOptionSecond+\noiseVarianced}{\symbolEnergy\constantPowerOfES\numberofconnectedES+2\noiseVarianced}~.
    \label{equation-conditionalproblem2}
\end{align}
and
\begin{align}
    \probability[{\signofdeltaElement[\indexCommunicationRound][{\indexES,\indexGradient}]=\SignofUsercentricElement[\indexED,\indexGradient][\indexCommunicationRound]|\numberOfEDsWithCorrectChoice=\numberOFEDsForOptionOne}]&=\frac{\meanOptionOneES}{\meanOptionOneES+\meanOptionTwoES}~,
    %\nonumber\\&\hspace{-14mm}=\frac{\symbolEnergy\constantPowerOfED\numberOFEDsForOptionOne+\noiseVariance}{\symbolEnergy\constantPowerOfED\numberofconnectedED+2\noiseVariance}~,
    \label{equation-conditionalproblem}
\end{align}
respectively. By using the definitions of  $\meanOptionOneES$ and $\meanOptionTwoES$ and substituting \eqref{equation-conditionalproblem} into \eqref{EScorrectdecision}, we obtain 
\begin{align}
    \correctDecisionED&=\sum^{\numberofconnectedED}_{\numberOFEDsForOptionOne=1}\frac{\symbolEnergy\constantPowerOfED\numberOFEDsForOptionOne+\noiseVariance}{\symbolEnergy\constantPowerOfED\numberofconnectedED+2\noiseVariance}\binom{\numberofconnectedED}{\numberOFEDsForOptionOne}\correctDecision[\indexGradient]^{\numberOFEDsForOptionOne}\incorrectDecision[\indexGradient]^{\numberofconnectedED-\numberOFEDsForOptionOne}\nonumber\\&=\frac{\symbolEnergy\constantPowerOfED\numberofconnectedED\correctDecision[\indexGradient]+\noiseVariance}{\symbolEnergy\constantPowerOfED\numberofconnectedED+2\noiseVariance}~.\label{eq:successClosedform}
\end{align}
By substituting \eqref{equation-conditionalproblem2} into \eqref{errorP} and using \eqref{eq:successClosedform}, we obtain $\probabilityIncorrect[\indexGradient,\indexED]$ as
\begin{align}
    \probabilityIncorrect[\indexGradient,\indexED]&=\sum^{\numberofconnectedES}_{\numberOFESsForOptionOne=1}\frac{\symbolEnergy\constantPowerOfED\numberOFESsForOptionSecond+\noiseVarianced}{\symbolEnergy\constantPowerOfED\numberofconnectedES+2\noiseVarianced}
    \binom{\numberofconnectedES}{\numberOFESsForOptionOne}{\correctDecisionED}^{\numberOFESsForOptionOne}\incorrectDecisionED^{\numberofconnectedES-\numberOFESsForOptionOne}\nonumber\\
    &\leq\frac{\noiseVarianced+\symbolEnergy\constantPowerOfES\numberofconnectedES\left({1-{\frac{\noiseVariance+\symbolEnergy\constantPowerOfED\numberofconnectedED\left(1-\frac{\sqrt{2}}{3S}\right)}{\symbolEnergy\constantPowerOfED\numberofconnectedED+2\noiseVariance}}}\right)}{\symbolEnergy\constantPowerOfES\numberofconnectedES+2\noiseVarianced}~,\label{errorPP}
\end{align}
for $S\triangleq{|\ElementsOfGradient[\indexED,\indexGradient][\indexCommunicationRound]|}/{\frac{\varianceBoundEle[\indexGradient]^2}{\sqrt{\batchSize}}}$. Accordingly, an upper bound for the stochasticity-induced error can be obtained as follows:
\begin{align}
    \sum^{\numberOfEdgeDevices}_{\indexED=1}\sum_{\indexGradient=1}^{\numberOfModelParametersED[]}|\ElementsOfGradient[\indexED,\indexGradient][\indexCommunicationRound]|\probabilityIncorrect[\indexED,\indexGradient]\leq A\norm{\localGradient[\indexED][\indexCommunicationRound]}_1+B\frac{\sqrt{2}}{3\sqrt{\batchSize}}\norm{\varianceBound}_1,
\end{align}
where $A$ and $B$ are defined in Theorem~\ref{th:main}.
%$A=\frac{\symbolEnergy\constantPowerOfED\numberofconnectedES+\noiseVarianced}{\symbolEnergy\constantPowerOfED\numberofconnectedES+2\noiseVarianced}-B$ and $B=\frac{\symbolEnergy\constantPowerOfED\numberofconnectedES\noiseVariance+{\symbolEnergy\constantPowerOfED}^2\numberofconnectedES\numberofconnectedED}{(\symbolEnergy\constantPowerOfED\numberofconnectedES+2\noiseVarianced)(\symbolEnergy\constantPowerOfED\numberofconnectedED+2\noiseVariance)}$.
By considering Assumption~1, an upper bound can then be obtained as follows:
\begin{align}
   &\lossFunctionGlobalMinimum-\lossFunctionLocal[\indexED][{\modelParametersAtIteration[0][\indexED]}]\nonumber\\&
   \leq\expectationOperator[{\sum^{\communicationRounds-1}_{\indexCommunicationRound=0} -\learningRate\numberOfEdgeDevices\norm{\localGradient[\indexED][\indexCommunicationRound]}_1+\frac{{\learningRate}^2}{2}\numberOfEdgeDevices\norm{\constantvector}_1+2\learningRate\sum^{\numberOfEdgeDevices}_{\indexED=1}\sum_{\indexGradient=1}^{\numberOfModelParametersED[]}|\ElementsOfGradient[\indexED,\indexGradient][\indexCommunicationRound]|\probabilityIncorrect[\indexED,\indexGradient]}][]\nonumber\\&
   %=\expectationOperator[{\sum^{\communicationRounds-1}_{\indexCommunicationRound=0} \learningRate\numberOfEdgeDevices\norm{\localGradient[\indexED][\indexCommunicationRound]}_1+\frac{{\learningRate}^2}{2}\numberOfEdgeDevices\norm{\constantvector}_1+2\learningRate\left( A\norm{\localGradient[\indexED][\indexCommunicationRound]}_1+B\frac{\sqrt{2}}{3\sqrt{\batchSize}}\norm{\varianceBound}_1\right)}][]\\\nonumber&
   =\expectationOperator[{\sum^{\communicationRounds-1}_{\indexCommunicationRound=0} -\learningRate(\numberOfEdgeDevices-2A)\norm{\localGradient[\indexED][\indexCommunicationRound]}_1+\frac{{\learningRate}^2}{2}\numberOfEdgeDevices\norm{\constantvector}_1+2\learningRate B\frac{\sqrt{2}}{3\sqrt{\batchSize}}\norm{\varianceBound}_1}][].\nonumber
%   \label{eq:finaleq}
\end{align}
Finally, by rearranging terms of the above equation
% \eqref{eq:finaleq}
and considering $\learningRate=\frac{1}{\communicationRounds}$ and $\batchSize=\frac{\communicationRounds}{\gamma}$, \eqref{eq:convergence} can be reached.
% \begin{align}
%     &\expectationOperator[\frac{1}{\communicationRounds}{\sum^{\communicationRounds-1}_{\indexCommunicationRound=0}\norm{\localGradient[\indexED][\indexCommunicationRound]}_1}][]\nonumber\\
%     %\nonumber&\leq\frac{1}{\learningRate(\numberOfEdgeDevices-2A){\sqrt{\communicationRounds}}}\left(\lossFunctionLocal[\indexED][{\modelParametersAtIteration[0][\indexED]}]-\lossFunctionGlobalMinimum+\frac{\communicationRounds{\learningRate}^2}{2}\numberOfEdgeDevices\norm{\constantvector}_1+2\communicationRounds\learningRate B\frac{\sqrt{2}}{3\sqrt{\batchSize}}\norm{\varianceBound}_1\right)\\
%     \nonumber&\leq\frac{1}{(\numberOfEdgeDevices-2A){\sqrt{\communicationRounds}}}\left(\lossFunctionLocal[\indexED][{\modelParametersAtIteration[0][\indexED]}]-\lossFunctionGlobalMinimum+\frac{1}{2}\numberOfEdgeDevices\norm{\constantvector}_1+2\sqrt{\gamma}B\frac{\sqrt{2}}{3}\norm{\varianceBound}_1\right).
% \end{align}
\end{IEEEproof}
%Let $\densityofpositiveEDnumber$ be a random variable for counting the number of \ac{ED}s with the correct decision, $\densityofpositiveEDnumber=$ 

%\subsection{Comparison with Other AirComp Schemes}

\section{Numerical Results}
\def\figuresize{1.75in}
\begin{figure}[t]
	\centering
	\subfloat[All classes, homogeneous data.]{\includegraphics[width =\figuresize]{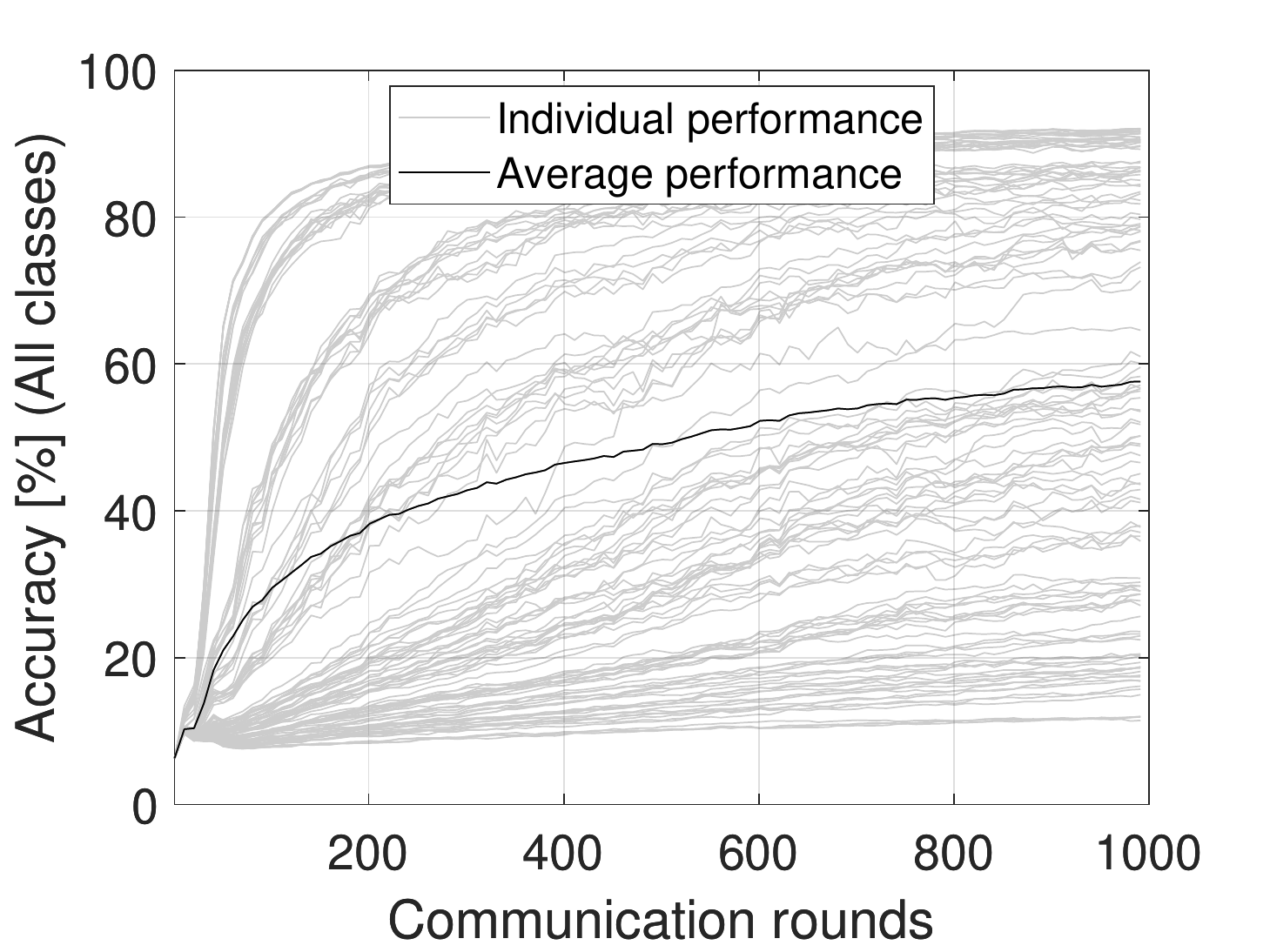}\label{subfig:acc_sc_20_acc_iid}}		
% 	\subfloat[All classes, heterogeneous data.]{\includegraphics[width =\figuresize]{figure_acc_dES_50_SNR_20_NxES_0_NxED_4_Areas_5_D_30000_thresAct_0_end.pdf}\label{subfig:acc_sc_20_acc_niid}}
    %\\
	%\subfloat[User-centric, homogeneous data.]{\includegraphics[width =\figuresize]{figure_accPer_dES_50_SNR_20_NxES_0_NxED_4_Areas_1_D_30000_thresAct_0_end.pdf}\label{subfig:accPer_sc_20_acc_iid}}		
	\subfloat[Personalized,  heterogeneous  data.]{\includegraphics[width =\figuresize]{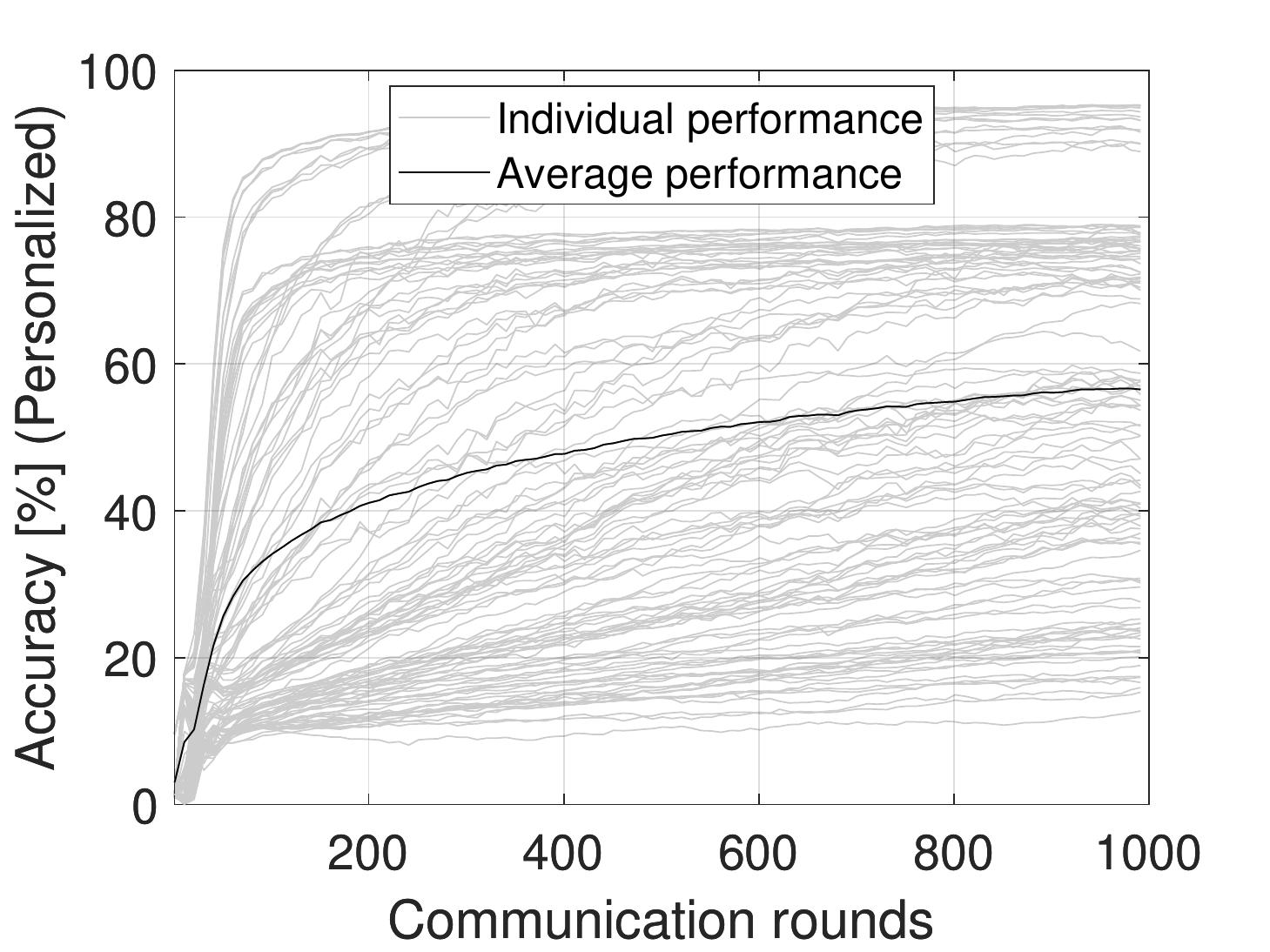}\label{subfig:accPer_sc_20_acc_niid}}
	\caption{Test accuracy - communication round for a single cell ($|\completeData|=30000$).}
	\label{fig:testAccSingle20dB}
	\centering
	\subfloat[All classes, homogeneous data.]{\includegraphics[width =\figuresize]{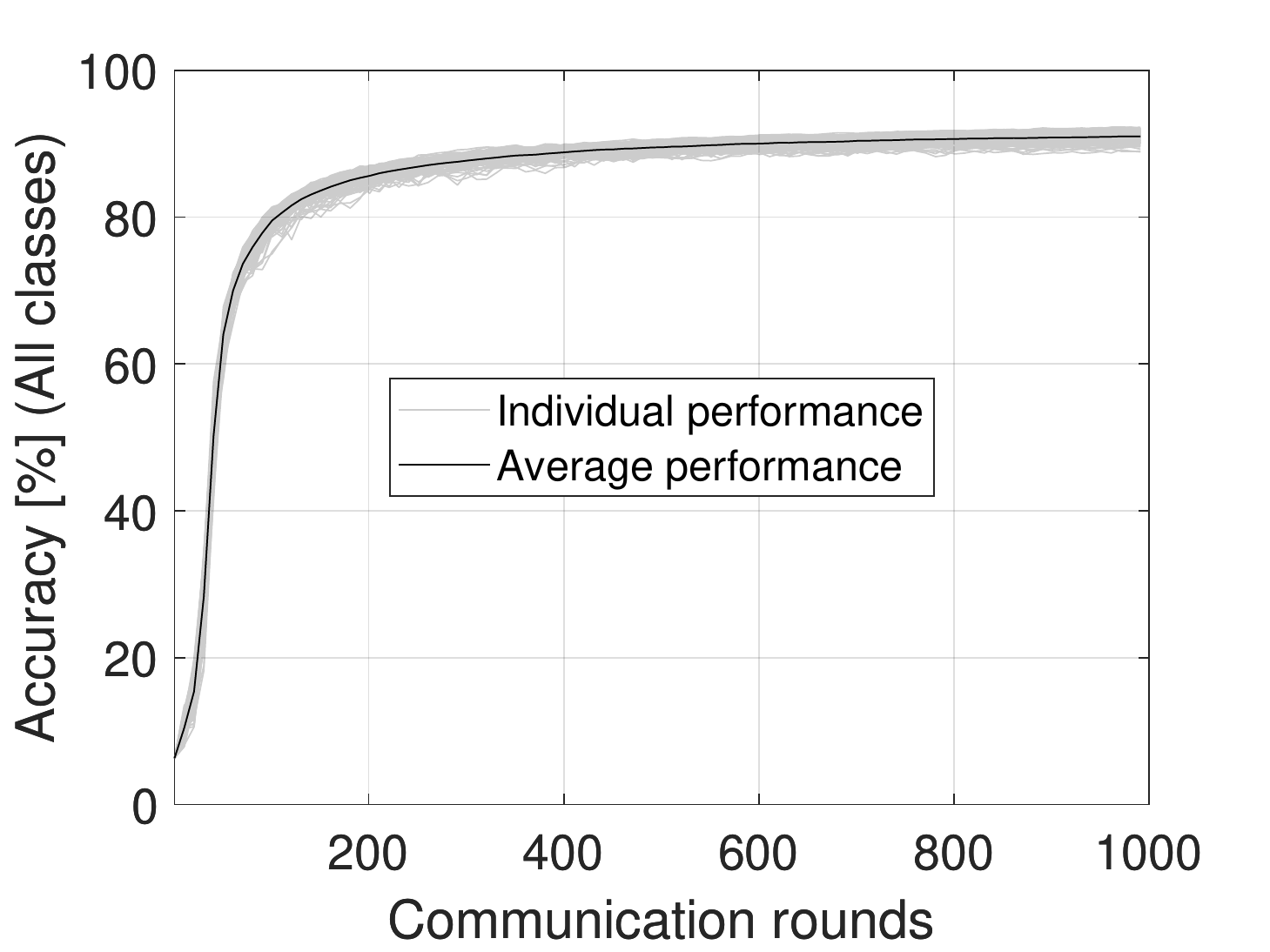}\label{subfig:acc_mc_20_acc_iid}}		
% 	\subfloat[All classes, heterogeneous data.]{\includegraphics[width =\figuresize]{figure_acc_dES_50_SNR_20_NxES_4_NxED_4_Areas_5_D_30000_thresAct_0_end.pdf}\label{subfig:acc_mc_20_acc_niid}}	
    %\\
	%\subfloat[User-centric, homogeneous data.]{\includegraphics[width =\figuresize]{figure_accPer_dES_50_SNR_20_NxES_4_NxED_4_Areas_1_D_30000_thresAct_0_end.pdf}\label{subfig:accPer_mc_20_acc_iid}}		
    \subfloat[Personalized,  heterogeneous  data.]{\includegraphics[width =\figuresize]{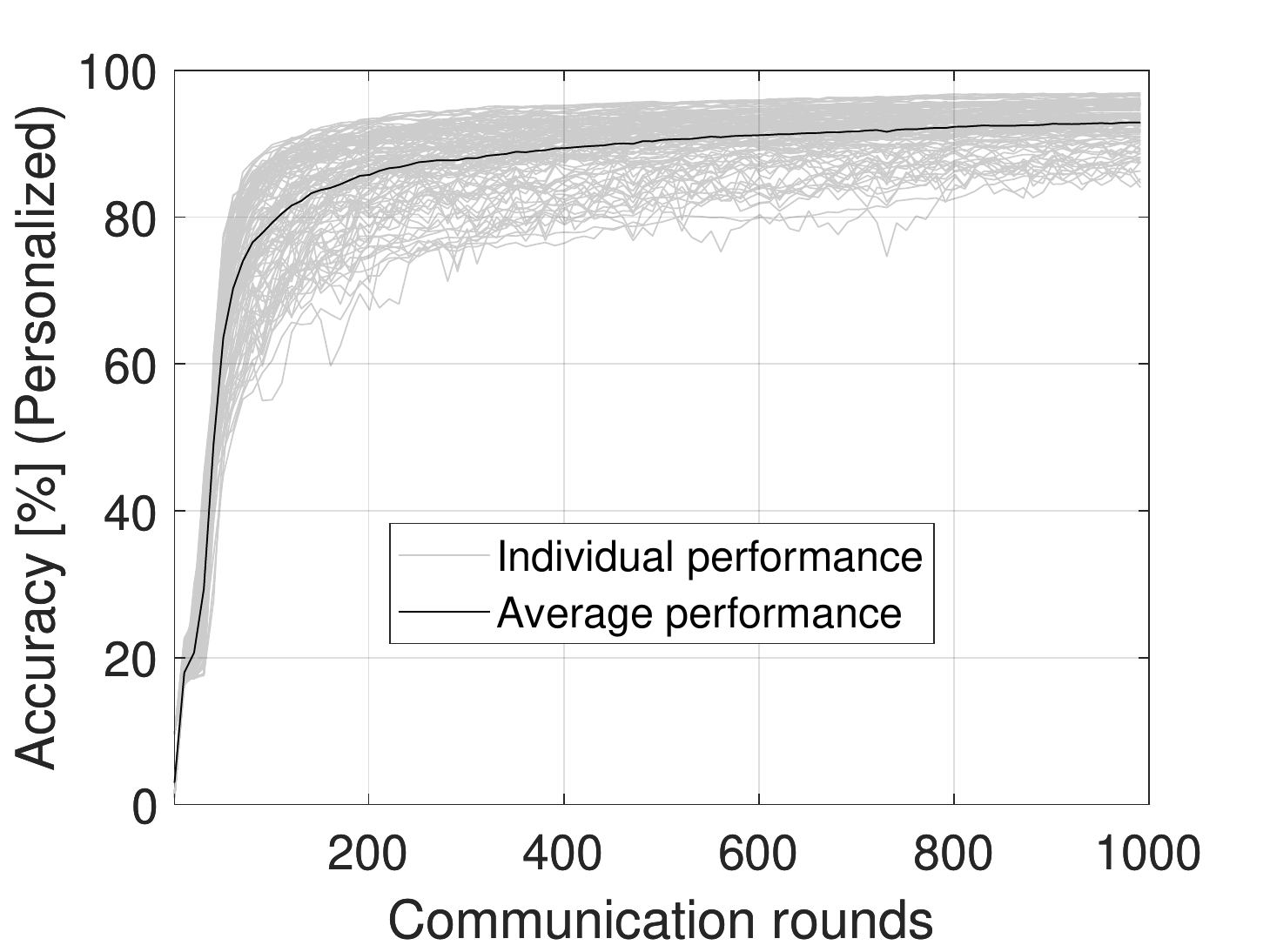}\label{subfig:accPer_mc_20_acc_niid}}		
	\caption{Test accuracy - communication round for multiple cells  ($|\completeData|=30000$).}
	\label{fig:testAccMulti20dB}
\end{figure}
To numerically evaluate multi-cell \ac{OAC}, we consider the learning task of handwritten-digit recognition over an  hexagonal  tessellation  with $77$ cells, i.e., $\numberOfEdgeServers=77$ \acp{ES}, where $\numberOfEdgeDevices=120$ \acp{ED}  are located at the cell edge and the distance between two adjacent \acp{ES} is $50$~meters (see \figurename~\ref{fig:testAccAreaSingle20dB}). Under this specific deployment, $\numberofconnectedED$ and  $\numberofconnectedES$ are approximately $6$ and $3$, respectively\footnote{We do not assume a fixed connectivity assumption for the numerical analysis. The received signal powers are governed by the path loss model.}. Our evaluation is limited to \ac{FSK-MV} since it is the only scheme that allows both \ac{OAC} in both \ac{UL} and \ac{DL}, to the best of our knowledge. 
%Hence,  we demonstrate the efficacy of multi-cell \ac{OAC} by comparing it with a single cell \ac{OAC} scenario.
For the large-scale channel model, we assume that the path loss exponent is $\pathlossExponent=4$ and the \ac{UL}  and  \ac{DL}  \acp{SNR} are set to $20$~dB for $\referenceDistanceUplink=\referenceDistanceDownlink=25/\cos(\pi/6)$.   %(i.e., an \ac{ES} receives the signals from all \acp{ES} regardless of the signal strength).
For the fading channel, we consider ITU Extended Pedestrian A (EPA) with no mobility in both \ac{UL} and \ac{DL}  and capture the long-term channel variations by regenerating the channels between the \acp{ES} and the \acp{ED} independently for each communication round. In this study, we also assume that the \ac{UL} and \ac{DL} channel realizations are independent of each other.
The subcarrier spacing and the \ac{CP} duration are set to $15$~kHz and $4.7$ $\mu$s, respectively. We use  $\numberOfActiveSubcarriers=1200$ subcarriers (i.e., the signal bandwidth is $18$~MHz). Therefore, $\syncError$ can be calculated as $55.6$ ns.

\begin{figure}[t]
	\centering
	\subfloat[All classes, homogeneous data.]{\includegraphics[width =\figuresize]{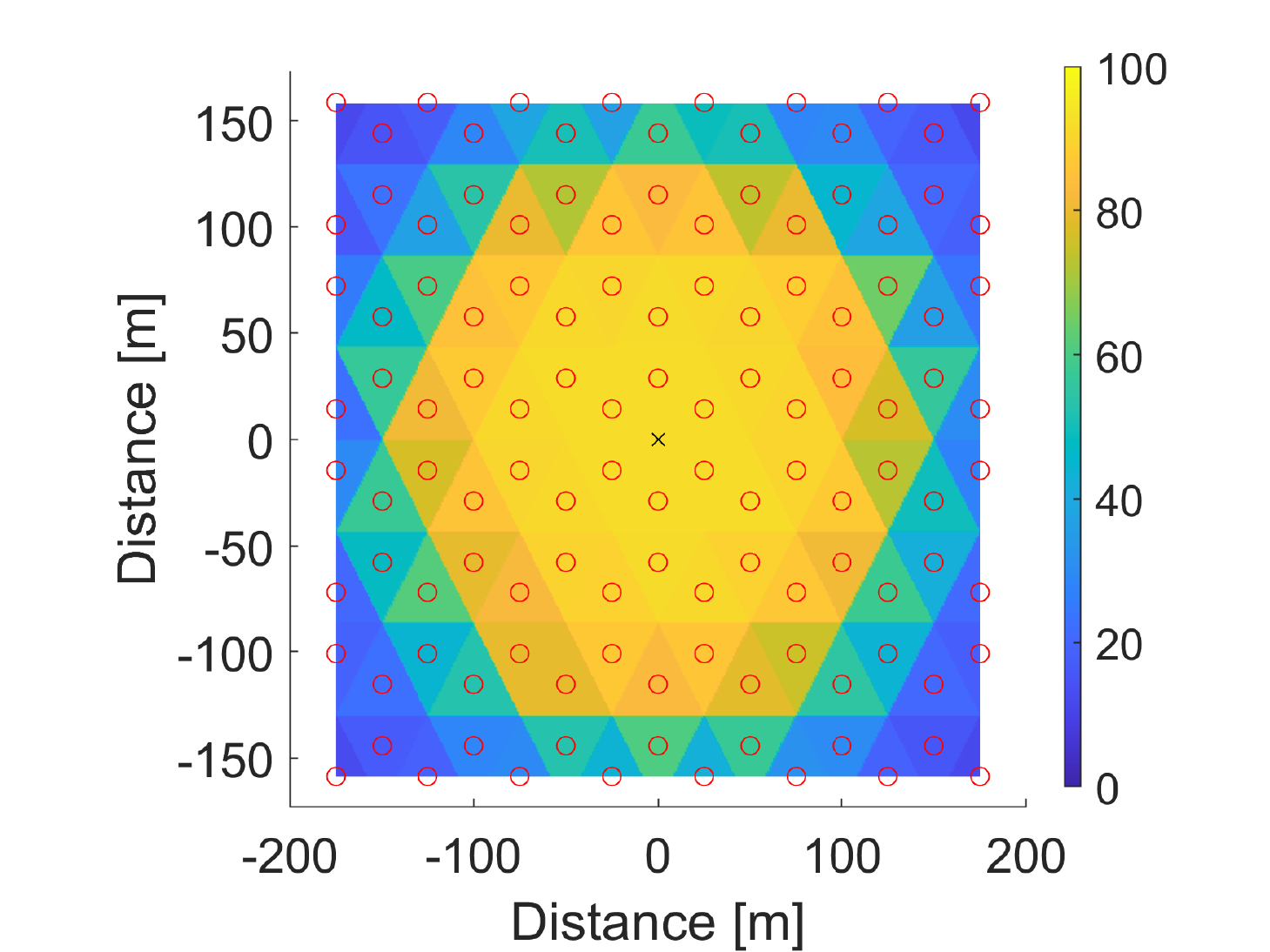}\label{subfig:dep_sc_20_acc_iid}}		
% 	\subfloat[All classes, heterogeneous  data.]{\includegraphics[width =\figuresize]{figure_dep_acc_dES_50_SNR_20_NxES_0_NxED_4_Areas_5_D_30000_thresAct_0_end.pdf}\label{subfig:dep_sc_20_acc_niid}}
    %\\
	%\subfloat[User-centric, homogeneous data.]{\includegraphics[width =\figuresize]{figure_dep_accPer_dES_50_SNR_20_NxES_0_NxED_4_Areas_1_D_30000_thresAct_0_end.pdf}\label{subfig:dep_sc_20_accPer_iid}}		
	\subfloat[Personalized, heterogeneous  data.]{\includegraphics[width =\figuresize]{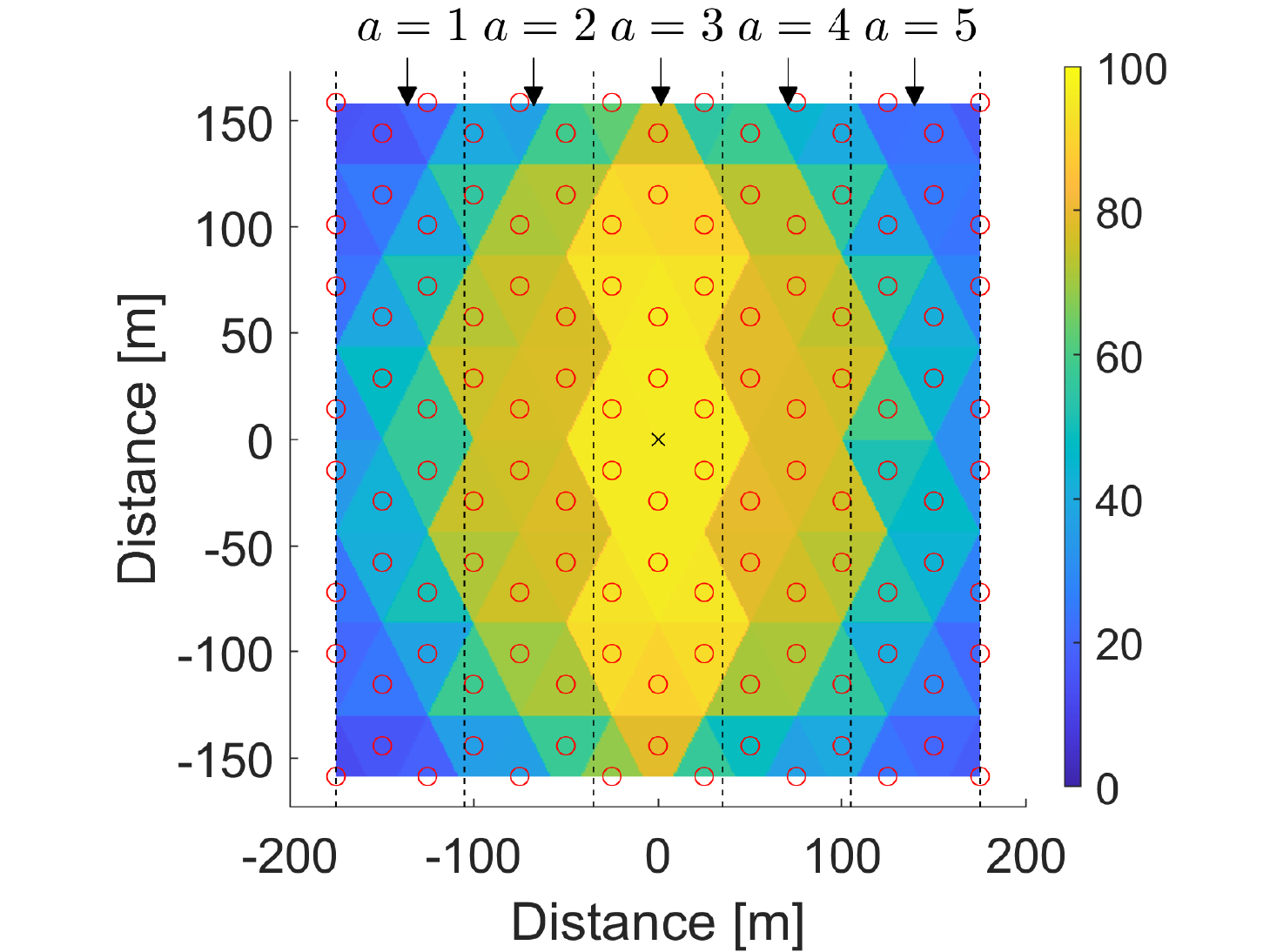}\label{subfig:dep_sc_20_accPer_niid}}	
	\caption{Distribution of the test accuracy for a single cell ($\times$: \ac{ES}, $\circ$: \ac{ED}, Blue: Low test accuracy, Yellow: High test accuracy, $|\completeData|=30000$).}
	\label{fig:testAccAreaSingle20dB}
	\centering
	\subfloat[All classes, homogeneous data.]{\includegraphics[width =\figuresize]{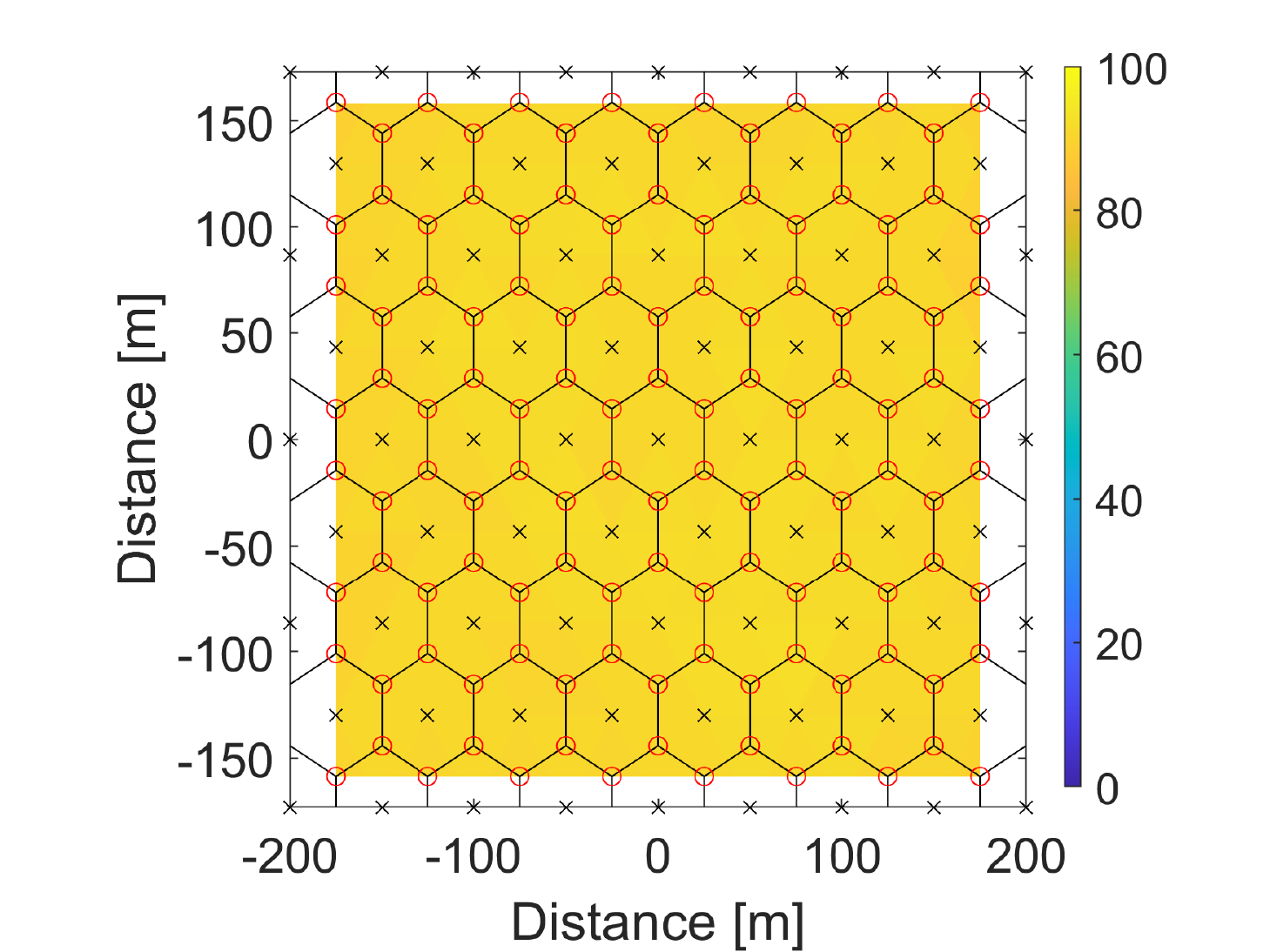}\label{subfig:dep_mc_20_acc_iid}}		
% 	\subfloat[All classes, heterogeneous  data.]{\includegraphics[width =\figuresize]{figure_dep_acc_dES_50_SNR_20_NxES_4_NxED_4_Areas_5_D_30000_thresAct_0_end.pdf}\label{subfig:dep_mc_20_acc_niid}}		
    %\\
	%\subfloat[User-centric, homogeneous data.]{\includegraphics[width =\figuresize]{figure_dep_accPer_dES_50_SNR_20_NxES_4_NxED_4_Areas_1_D_30000_thresAct_0_end.pdf}\label{subfig:dep_mc_20_accPer_iid}}		
    \subfloat[Personalized, heterogeneous  data.]{\includegraphics[width =\figuresize]{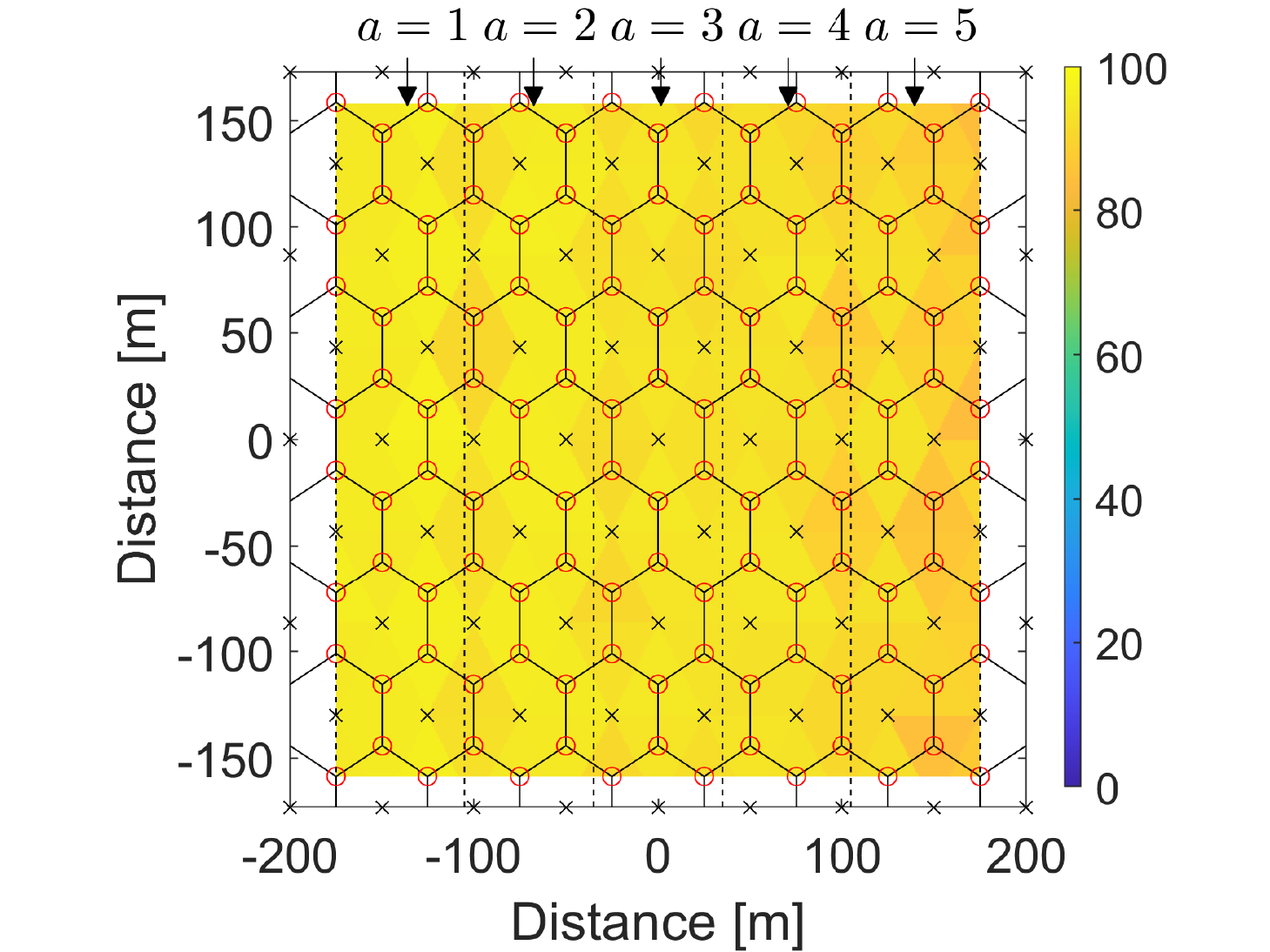}\label{subfig:dep_mc_20_accPer_niid}}		
	\caption{Distribution of the test accuracy for multiple cells ($\times$: \ac{ES}, $\circ$: \ac{ED}, Blue: Low test accuracy, Yellow: High test accuracy,   $|\completeData|=30000$).}
	\label{fig:testAccAreaMulti20dB}
	\centering
	\subfloat[All classes, homogeneous data.]{\includegraphics[width =\figuresize]{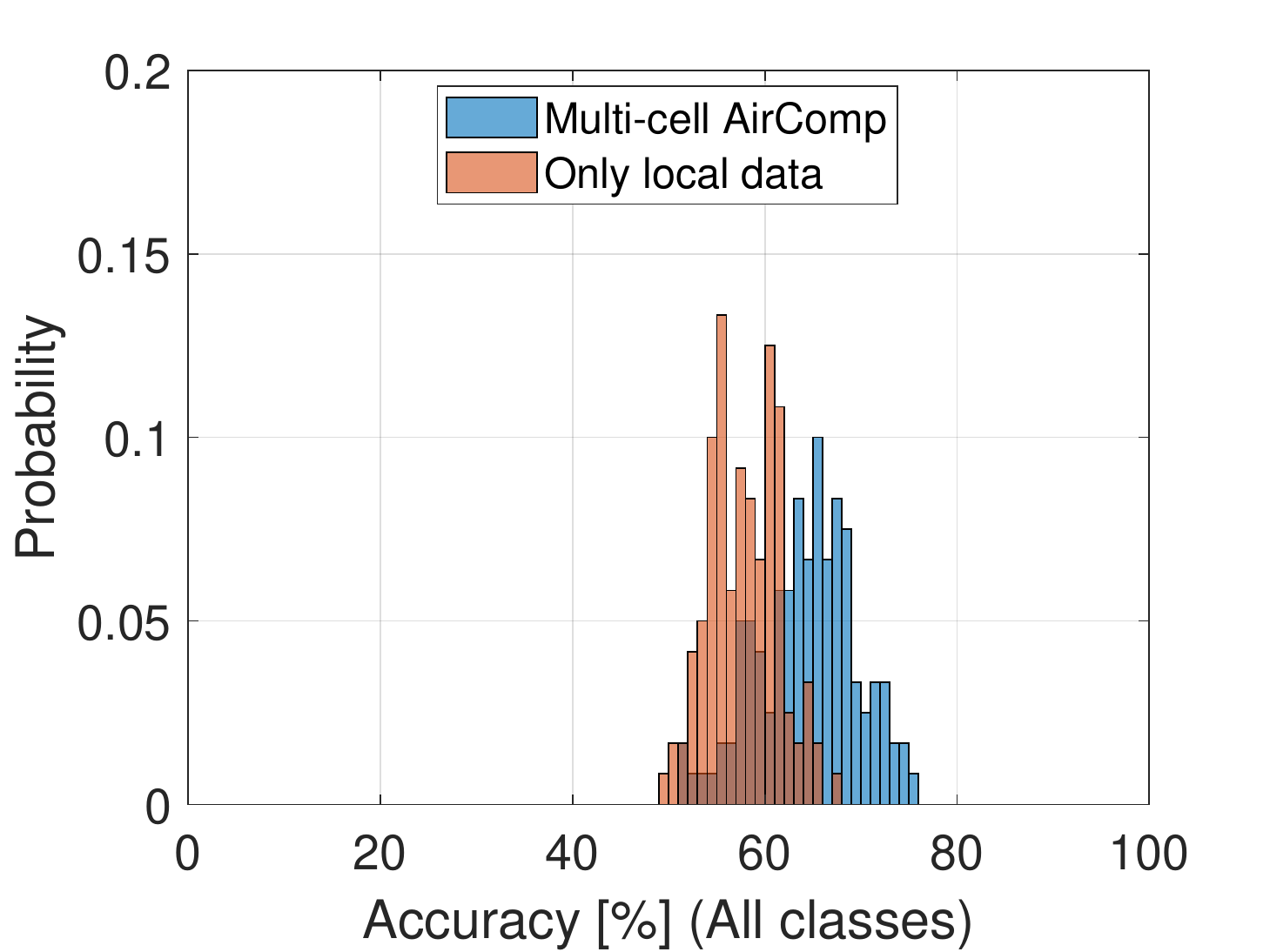}\label{subfig:histdep_mc_20_acc_iid}}		
% 	\subfloat[All classes, heterogeneous  data.]{\includegraphics[width =\figuresize]{figure_histAcc_dES_50_SNR_20_NxES_0_NxED_4_Areas_5_thresAct_0_end.pdf}\label{subfig:histdep_mc_20_acc_niid}}		
    \subfloat[Personalized, heterogeneous  data.]{\includegraphics[width =\figuresize]{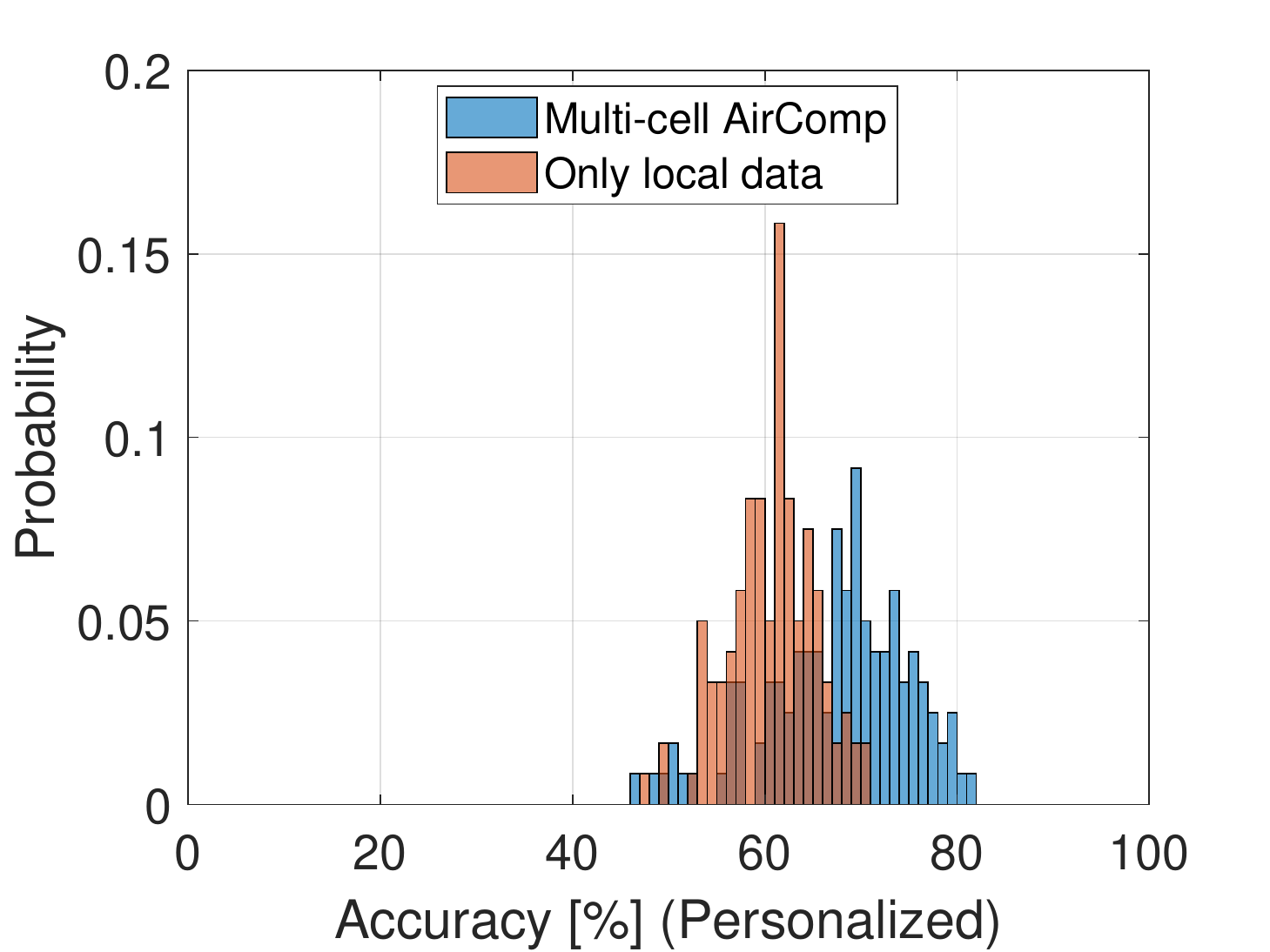}\label{subfig:histdep_mc_20_accPer_niid}}		
	\caption{Test accuracy with multi-cell \ac{FL} with the proposed \ac{OAC} or the training based on only local data after 400 iterations  ($|\completeData|=5000$).}
	\label{fig:testAcchistMulti20dB}
\end{figure}

For the local data at the \acp{ED}, we use the MNIST database that contains labeled handwritten-digit images size of $28\times28$ from digit 0 to digit 9. We consider both homogeneous data and heterogeneous data distribution in the cell. To prepare the data, we first choose $|\completeData|\in\{5000,30000\}$ training images from the database, where each digit has the identical number of images.  
For the scenario with the homogeneous data distribution, each local dataset has approximately an equal number of distinct images for each digit. For the scenario with the heterogeneous data distribution,  we assume that the distribution of the images depends on the locations of the \acp{ED}. To this end, we divide the area into 5 identical parallel areas, where the \acp{ED} located in the $\indexArea$th area have the data samples with the labels $\{\indexArea-1,\indexArea,1+\indexArea,2+\indexArea,3+\indexArea,4+\indexArea\}$ for $\indexArea\in\{1,\mydots,5\}$  (see \figurename~\ref{fig:testAccAreaSingle20dB}\subref{subfig:dep_mc_20_accPer_niid}).  Hence, the availability of the labels gradually changes. %The difference between the homogeneous and heterogeneous data distribution is illustrated in \figurename~\ref{fig:niid}. 
The model at \acp{ED} is based on a \ac{CNN} described in \cite{sahinCommnet_2021}. 
%that includes one $5\times5$ and two $3\times3$ convolutional layers, where each of them is followed by a  batch normalization layer and \ac{ReLU} activation follow each of them. All convolutional layers have $20$ filters. After the third \ac{ReLU}, a fully-connected layer with 10 units and a softmax layer are utilized. At the input layer, no normalization is applied. 
It has $\numberOfModelParametersED[]=123090$ learnable parameters, which corresponds to $\numberOfOFDMSymbols=206$ OFDM symbols in both \ac{UL} and \ac{DL}, respectively. The learning rate is  $0.0001$. The batch size $\batchSize$ is $16$. For the test accuracy calculation, we use $10000$ test samples available in the MNIST database. For the personalized test accuracy, we test the models based on only the classes available at the \ac{ED}'s local dataset.

In \figurename~\ref{fig:testAccSingle20dB}, we evaluate the test accuracy versus communication round in a single cell under homogeneous and heterogeneous data distributions. When there is only a single \ac{ES} for the aggregation and the data distribution in the area is homogeneous, only a few number of \acp{ED} obtain a high test accuracy, while a majority of \acp{ED} fails to recognize the digits as shown in \figurename~\ref{fig:testAccSingle20dB}\subref{subfig:acc_sc_20_acc_iid}. %The best performance degrades when the data distribution is heterogeneous as can be seen in \figurename~\ref{fig:testAccSingle20dB}\subref{subfig:acc_sc_20_acc_niid}. 
The personalized test accuracy results for heterogeneous data distribution in \figurename~\ref{fig:testAccSingle20dB}\subref{subfig:accPer_sc_20_acc_niid} are also low (i.e., the \acp{ED} cannot even learn the classes that are available at their local datasets). In \figurename~\ref{fig:testAccMulti20dB}, we consider multi-cell scenario. When the data distribution is homogeneous, all \acp{ED} result in higher test accuracy results
as demonstrated in \figurename~\ref{fig:testAccMulti20dB}\subref{subfig:acc_mc_20_acc_iid}. %However, in the case of heterogeneous data distribution, the test accuracy based on all classes is around 50\% as shown in \figurename~\ref{fig:testAccMulti20dB}\subref{subfig:acc_mc_20_acc_niid}. 
The personalized test accuracy is also high for the heterogeneous data distribution as can be seen in \figurename~\ref{fig:testAccMulti20dB}\subref{subfig:accPer_mc_20_acc_niid}. This demonstrates that
%that although the \acp{ED} do not learn to recognize the digits that are not available on its dataset (or the ones at the neighboring \acp{ED}), 
 \acp{ED} learn to classify the labels while being harmonious with other \acp{ED} in the wireless network with the proposed \ac{OAC} framework. 
\figurename~\ref{fig:testAccMulti20dB} shows that the convergence for this specific learning task can be achieved approximately after $200$  rounds. Thus, the amount of consumed time-frequency resources  can be calculated as $2\times(66.7+4.7)\mu s\times 206 \times 200=5.88$ seconds over $18$~MHz, respectively.

In \figurename~\ref{fig:testAccAreaSingle20dB}\subref{subfig:dep_sc_20_acc_iid} and \figurename~\ref{fig:testAccAreaSingle20dB}\subref{subfig:dep_sc_20_accPer_niid}, we show the distribution of the test accuracy in the area. The single-cell \ac{OAC} suffers from path loss: The far \acp{ED}' votes cannot contribute the \ac{MV} decision in the \ac{UL}. Similarly, the \ac{ES}'s signal is not strong at the far \acp{ED} in the \ac{DL}. Therefore, only nearby \acp{ED} get benefit from the \ac{FEEL} and the ones have similar data distribution. On the other hand, multi-cell \ac{OAC} yields almost a uniform distribution for both homogeneous and heterogeneous data as it can be seen in \figurename~\ref{fig:testAccAreaMulti20dB}\subref{subfig:dep_mc_20_acc_iid} and \figurename~\ref{fig:testAccAreaMulti20dB}\subref{subfig:dep_mc_20_accPer_niid}, respectively.

In \figurename~\ref{fig:testAcchistMulti20dB}, we  evaluate if the proposed \ac{OAC} method is superior to the case where each \ac{ED} performs the training based on its own local data. To this end,  we intentionally reduce $|\completeData|$ to $5000$ and set $\learningRate=0.01$ to demonstrate if the \acp{ED} are able to leverage the data at the neighbouring \acp{ED} through \ac{FEEL}. We plot the histogram of the test accuracy after $400$ iterations for both cases. The results show that, in both homogeneous and heterogeneous data distributions, the proposed concept improves the average test accuracy based on all classes and personalized test accuracy in this scenario. % and improves the overall learning experience 

\section{Concluding Remarks}
In this paper, we present a multi-cell \ac{OAC} framework where the aggregations occur  in both \ac{UL} and \ac{DL} across multiple cells through a non-coherent \ac{OAC} scheme, i.e., \ac{FSK-MV}. We also prove the convergence of \ac{FEEL} under a fixed-connectivity assumption. Finally, we evaluate the test accuracy of the multi-cell \ac{OAC} by comparing it with the one for a single-cell scenario for homogeneous and heterogeneous data distributions. Our numerical results show that %in a multi-cell scenario, \acp{ED} can classify the labels without availability of some digits on their datasets, which 
the proposed approach is a promising solution to achieve a high-test accuracy at the \acp{ED} by exploiting the interference among multiple cells. In this study, our analysis is based on regular tessellation. 
For an irregular deployment, the interference distributions in \ac{UL} and \ac{DL} need to be considered for the convergence analysis, which will be investigated in future work.
%a higher test accuracy and superiority of this framework compared to a single-cell framework.

\acresetall
\bibliographystyle{IEEEtran}
\bibliography{references}

\end{document}